\documentclass[12pt,english]{article}
\usepackage[a4paper,
            bindingoffset=0.2in,
            left=1in,
            right=1in,
            top=1in,
            bottom=1in,
            footskip=.25in]{geometry}

\usepackage[doublespacing]{setspace}
\usepackage{natbib}
\usepackage{booktabs}
\usepackage[nomarkers, figuresonly, nofiglist]{endfloat}
\usepackage{inputenc}
\usepackage{graphicx}
\usepackage{tikz}
\usepackage{dblfloatfix}
\usepackage{multirow}
\usepackage{rotating}
\usepackage{lscape} 
\usepackage{longtable}
\usepackage[T1]{fontenc}
\usepackage{pgfplots}
\usepackage{afterpage}
\usepackage{url}

\usepackage{caption}
\usepackage{subcaption}
\usepackage{hyperref}
\usepackage{timet}
\usepackage{color}
\usepackage[draft]{todonotes}   
\usepackage[final, authormarkup=none]{changes}
\usepackage{comment}

\definechangesauthor[name={Andreas}, color={green}]{ar}
\definechangesauthor[name={Mike}, color={red}]{ml}

\begin{document} 

\begin{center}
    \textbf{\LARGE Bayesian regional moment tensor from ocean bottom seismograms recorded in the Lesser Antilles: Implications for regional stress field}\\
    \vspace{0.4cm}
    \textbf{\small 
    Mike Lindner $^1$,  
    Andreas Rietbrock $^1$, 
    Lidong Bie $^2$,
    Saskia Goes $^3$,
    Jenny Collier $^3$,
    Catherine Rychert $^4$,
    Nicholas Harmon $^4$,
    Stephen P. Hicks $^3$
    Tim Henstock $^4$
    and the VoiLA working group\footnote[5]{See full list of VoiLA working group members at \url{http://www.voila.ac.uk/index.php/project_participants/}}}\\
    \vspace{0.2cm}
    \textbf{\small 
    $^1$ Karlsruhe Institute of Technology, 
    $^2$ University of East Anglia
    $^3$ Imperial College London,
    $^4$ University of Southampton
    }\\
\end{center}


\section*{Abstract}
Seismic activity in the Lesser Antilles (LA) is characterized by strong regional variability along the arc reflecting the complex subduction setting and history. Although routine seismicity monitoring can rely on an increasing number of island stations, the island-arc setting means that high-resolution monitoring and detailed studies of fault structures require a network of ocean bottom seismometers (OBS). As part of the 2016-2017 \textbf{Vo}lat\textbf{i}le recycling at the \textbf{L}esser \textbf{A}ntilles arc (VoiLA) project, we deployed 34 OBS stations in the fore- and back-arc. During the deployment time, 381 events were recorded within the subduction zone. In this paper, we perform full-waveform regional moment tensor (RMT) inversions, to gain insight into the stress distribution along the arc and at depth. We developed a novel inversion approach, Am$\Phi$B - “Amphibious Bayesian”, taking into account uncertainties associated with OBS deployments. Particularly, the orientation of horizontal components (alignment uncertainty) and the high noise level on them due to ocean microseisms are accounted for using Am$\Phi$B. The inversion is conducted using a direct, uniform importance sampling of the fault parameters within a multi-dimensional tree structure: the uniXtree-sampling algorithm. We show that the alignment of the horizontal OBS components, particularly in high noise level marine environments, influences the obtained source mechanism when using standard least-squares (L2) RMT inversion schemes, resulting in systematic errors in the recovered focal mechanisms including high artificial compensated linear vector dipole (CLVD) contributions. Our Bayesian formulation in Am$\Phi$B  reduces these CLVD components by nearly 60\% and the aberration of the focal geometry as measured by the Kagan angle by around 40\% relative to a standard L2 inversion. Subsequently, we use Am$\Phi$B-RMT to obtain 45 (Mw > 3.8) regional MT solutions, out of which 39 are new to any existing database. Combining our new results with existing solutions, we subsequently analyze a total of 151 solutions in a focal mechanism classification (FMC) diagram and map them to the regional tectonic setting. We also use our newly compiled RMT database to perform stress tensor inversions along the LA subduction zone. On the plate interface, we observe the typical compressional stress regime of a subduction zone and find evidence for upper-plate strike slip and normal fault behaviour in the north that becomes a near arc-perpendicular extensional stress regime towards the south. A dominant slab perpendicular extensional stress regime is found in the slab at  100-200 km beneath the central part of the arc. We interpret this stress condition to be a result of  slab pull varying along the arc due to partial slab detachment along previously hypothesized lateral slab tear near Grenada, at the southern end of the LA arc,  leading to reactivation of preexisting structures around the subducted Proto-Caribbean ridge.

\section{Introduction}
\begin{figure}
\centering
\includegraphics[width=1.0\linewidth]{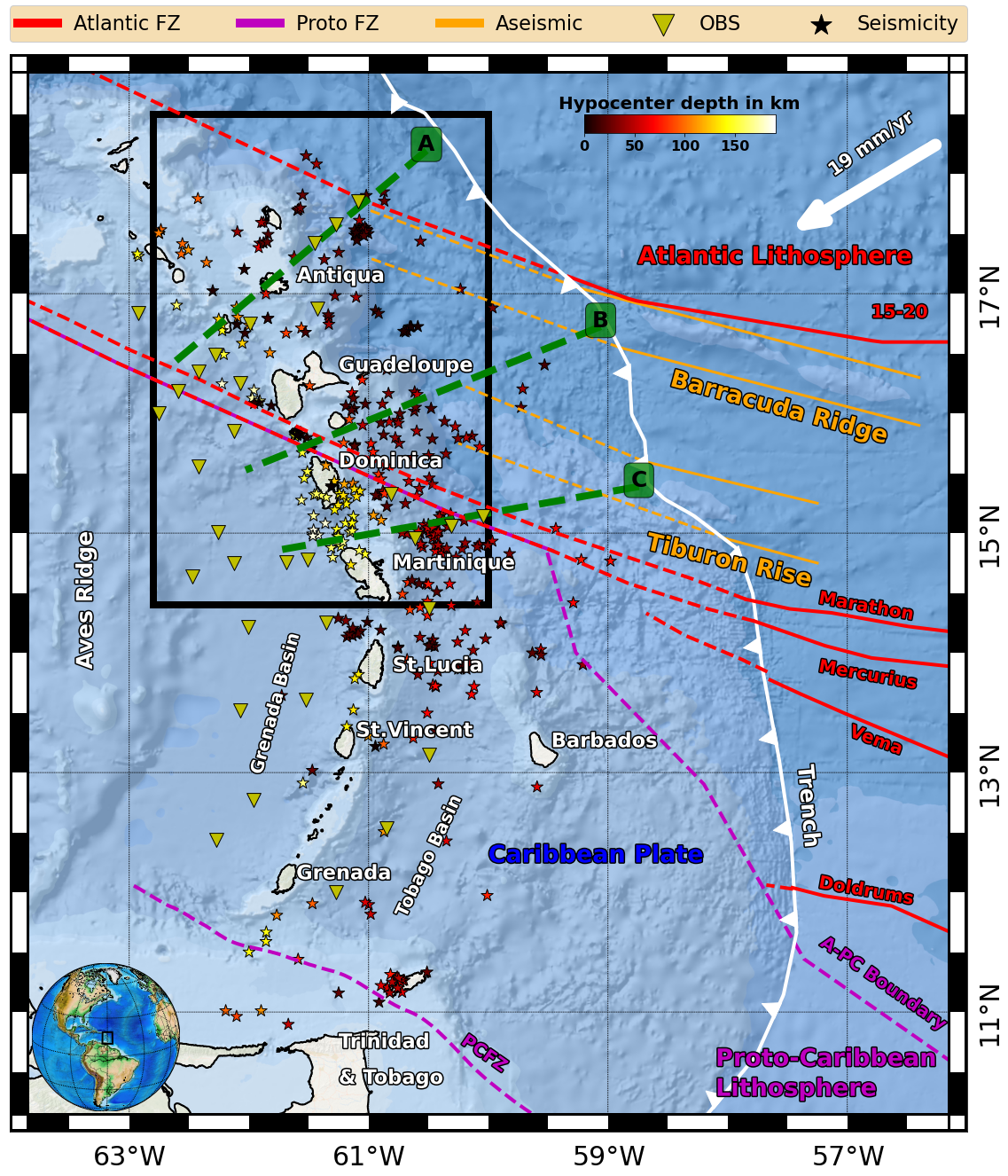}
\caption{\href{http://server.arcgisonline.com/arcgis/rest/services/Ocean/World_Ocean_Base/MapServer}{Bathymetric map} of the Lesser Antilles subduction zone. The black box outlines the target area of this study.  Triangles show the locations of the VoiLA OBS network, consisting of 32 broadband stations, within the fore- and back-arc region. Prominent fracture zones on the Atlantic lithosphere are marked in red, aseismics in orange and inferred features of the Proto-Caribbean lithosphere are marked in magenta. The dashed lines show the projected positions of these features along the subducted slab. Three cross-sections (A-C) are marked as green dotted lines. Background seismicity (black stars) during the operational time is taken from \citep{Bie2019}.}
\label{fig:Map1}
\end{figure}

\begin{figure}
     \centering
     \begin{subfigure}[b]{0.3\textwidth}
         \centering
         \includegraphics[width=\textwidth]{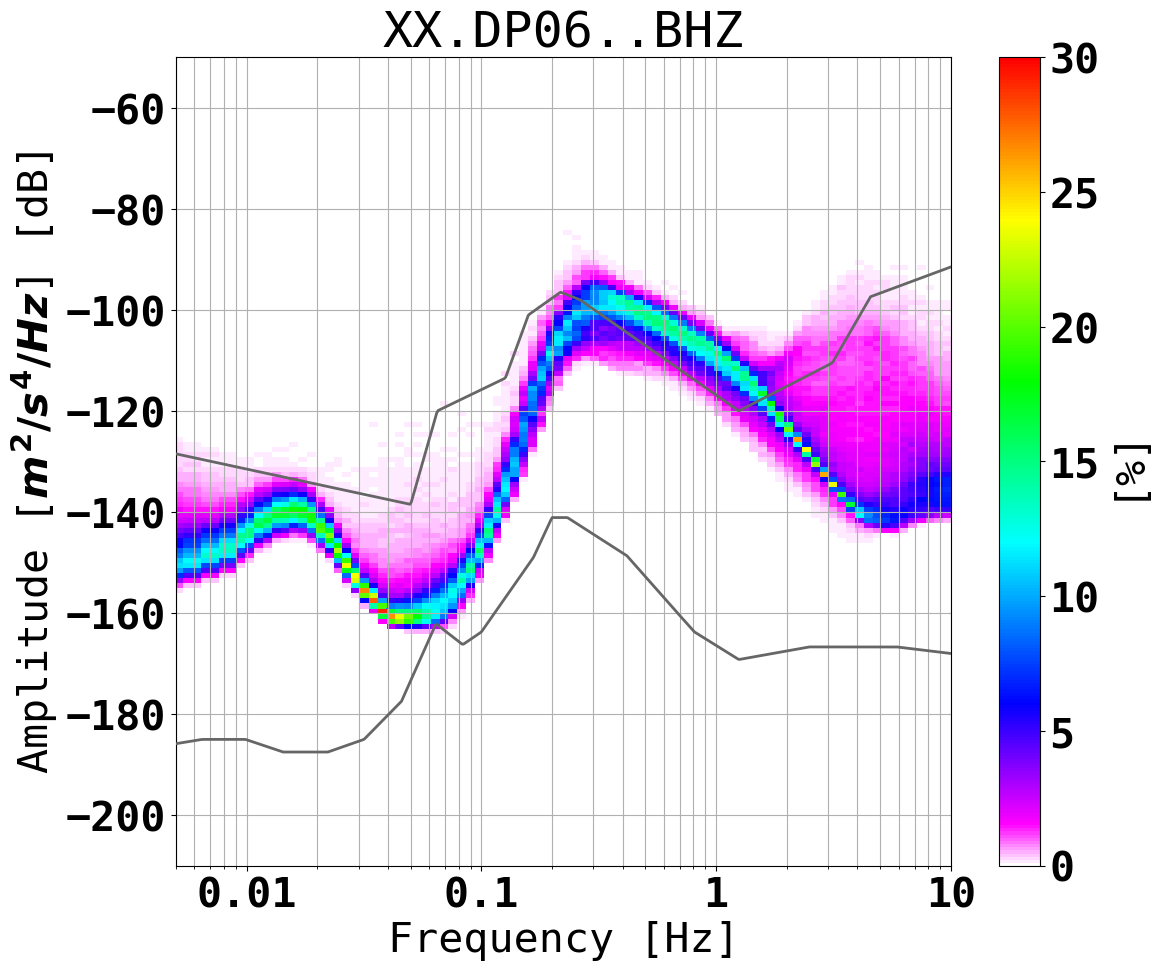}
         \caption{}
         \label{fig:PSD_Z}
     \end{subfigure}
     \hfill
     \begin{subfigure}[b]{0.3\textwidth}
         \centering
         \includegraphics[width=\textwidth]{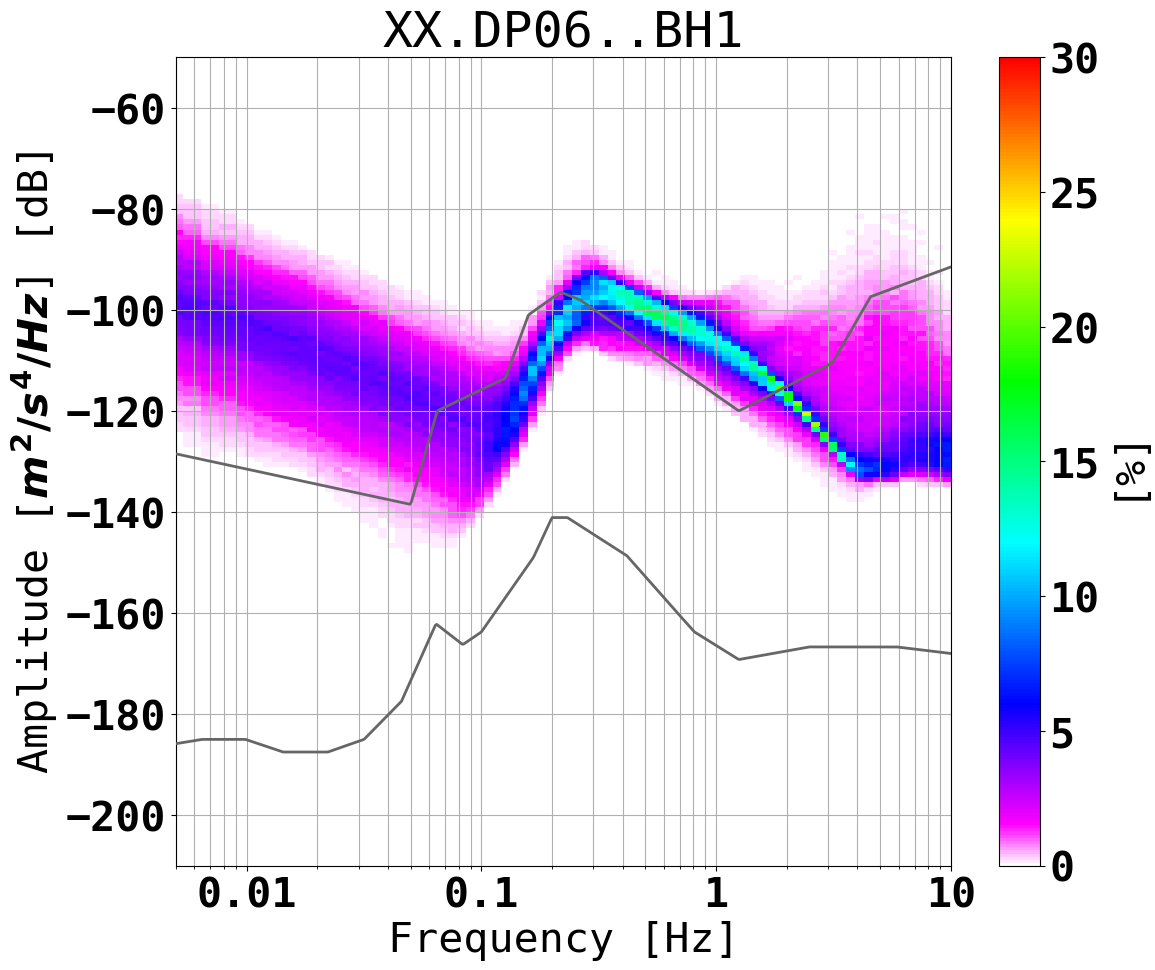}
         \caption{}
         \label{fig:PSD_1}
     \end{subfigure}
     \hfill
     \begin{subfigure}[b]{0.3\textwidth}
         \centering
         \includegraphics[width=\textwidth]{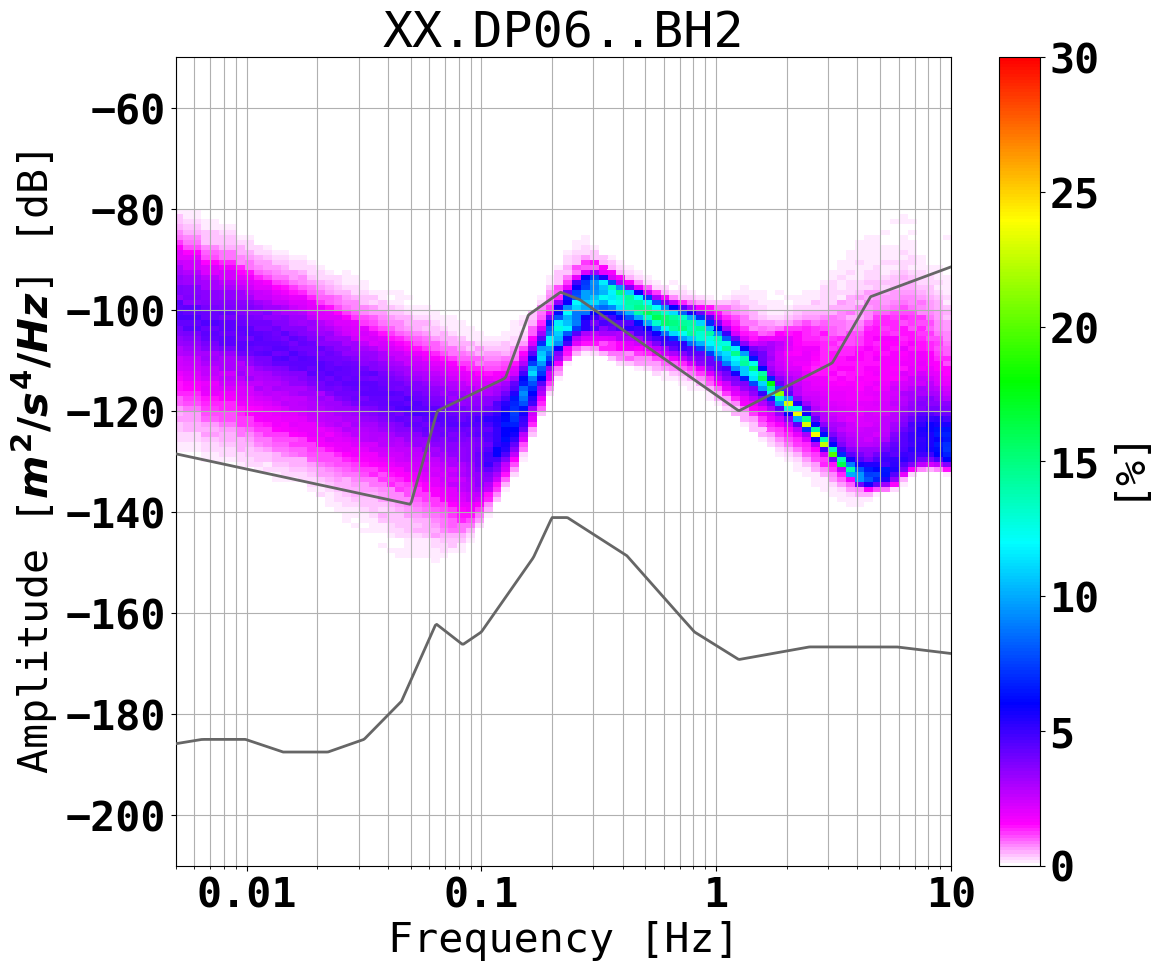}
         \caption{}
         \label{fig:PSD_2}
     \end{subfigure}\\
     \begin{subfigure}[b]{0.3\textwidth}
         \centering
         \includegraphics[width=\textwidth]{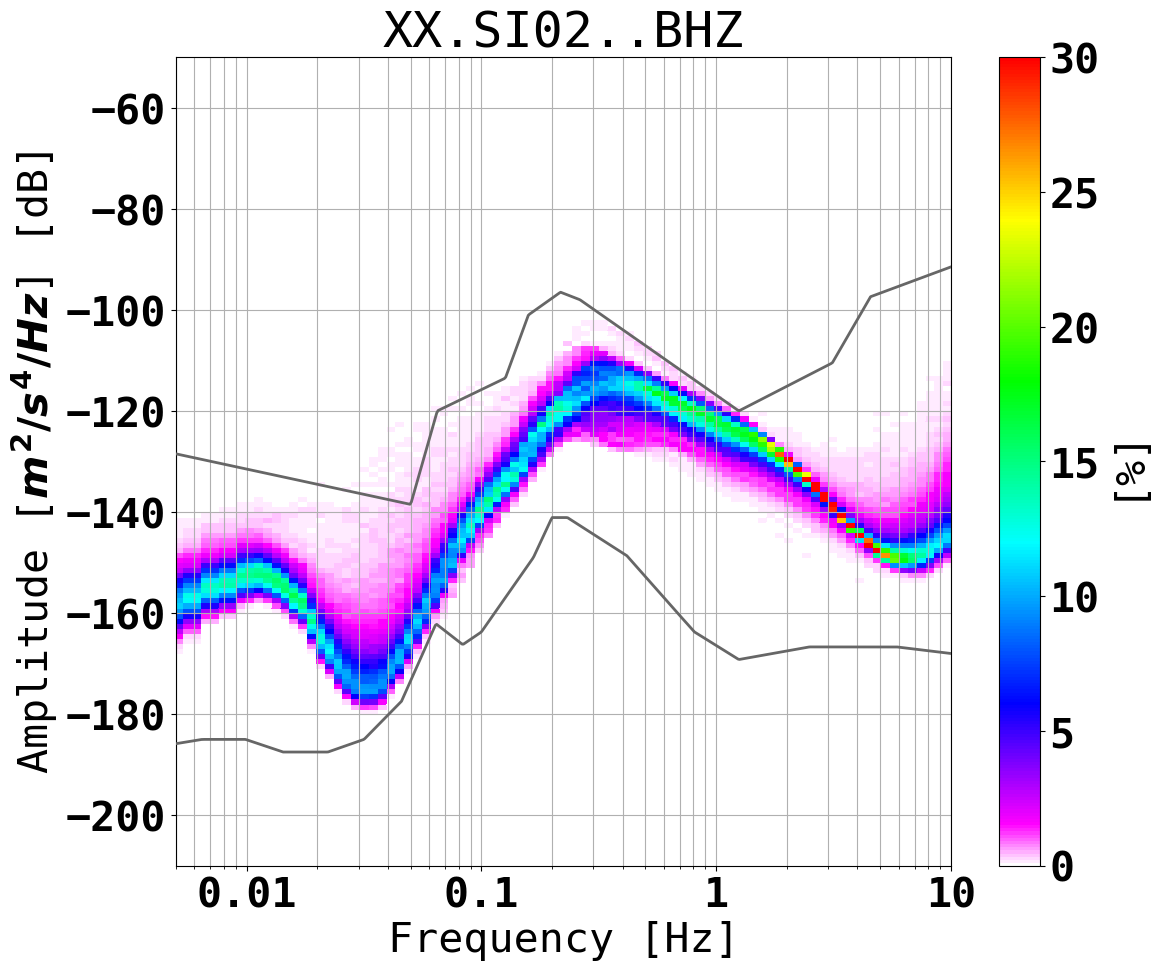}
         \caption{}
         \label{fig:PSD_Z}
     \end{subfigure}
     \hfill
     \begin{subfigure}[b]{0.3\textwidth}
         \centering
         \includegraphics[width=\textwidth]{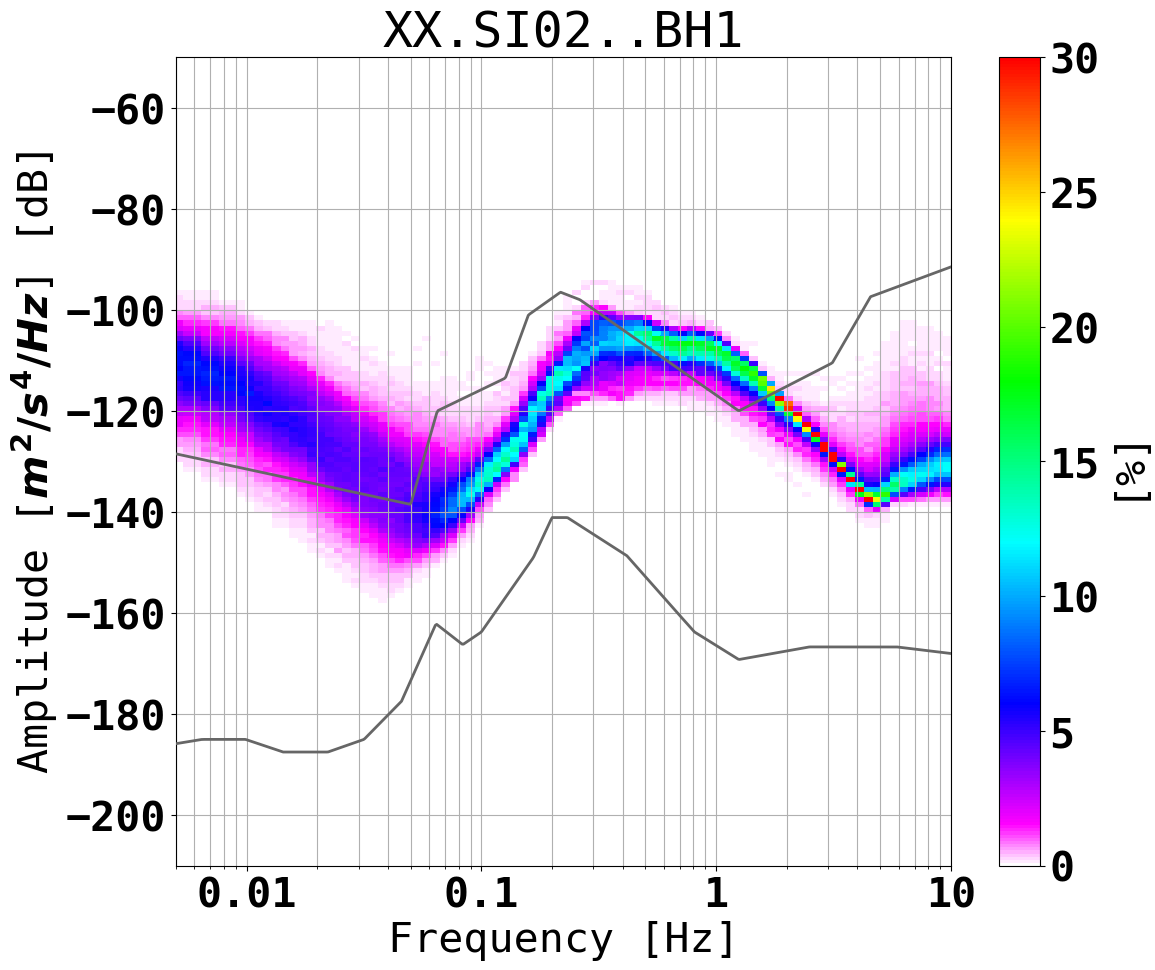}
         \caption{}
         \label{fig:PSD_1}
     \end{subfigure}
     \hfill
     \begin{subfigure}[b]{0.3\textwidth}
         \centering
         \includegraphics[width=\textwidth]{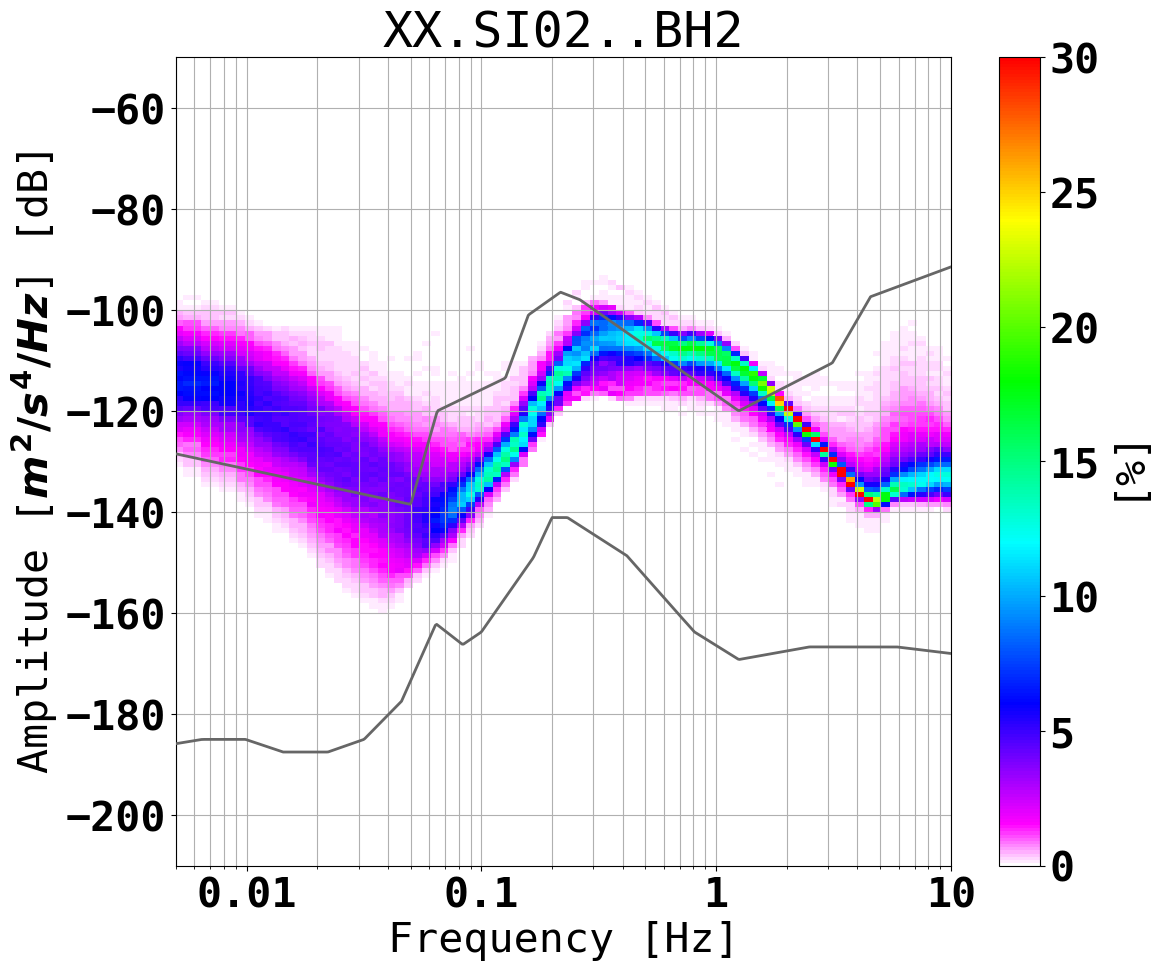}
         \caption{}
         \label{fig:PSD_2}
     \end{subfigure}
\caption{Probabilistic Power Spectral Density (PPSD) at stations DP06 (a-c) and SI02 (d-f) computed for a total of 250 days of continuous data records in 2016.}
\label{fig:PSD}
\end{figure}

The Lesser Antilles (LA) island arc (Fig. \ref{fig:Map1}), located along the eastern margin of the Caribbean Sea, is part of a small yet highly-complex subduction zone system that is driven by south-westward motions of the conjoined North and South American plates \citep{Bouysse1990, Harris2018, Allen2019}. Regional seismicity strongly varies along the arc \citep{Hayes2013, Schlaphorst2016, Bie2019}. The incoming oceanic plate is marked by large bathymetric structures north of 15$^\circ$N that significantly affect the activity within the shallow fore-arc. These prominent structures are the Barracuda Ridge entering the trench at around $16.75^\circ$N and the Tiburon Rise around $15.25^\circ$N \citep{Laigle2013b}. Earthquake activity follows the outline of the arc and reaches down to a maximum depth of $\sim$190 km around the island of Martinique. To the north of Martinique at $\sim$14.5°N, the seismicity rate is highest, especially for events with a magnitude greater than 5, whereas to the south, between St Lucia and  Grenada, an area with a much lower seismicity rate can be found. Even further south, seismic activity increases again near Tobago.
\newline
\noindent
Recent studies (e.g. \citep{Ruiz2013,Laigle2013,Laigle2013b,GONZALEZ2017214,Paulatto2017}) focus mostly on the seismically-active northern part of the LA arc between $14.5^\circ$ and $18.5^\circ$N. The main structural features of the subducted slab in this part of the arc include three prominent fracture zones i.e., the 15-20, Marathon, and Mercurius (e.g. \citep{Harmon2019}) and a domain boundary that separates plate material formed along the Proto-Caribbean and Equatorial Atlantic ridges \citep{Cooper2020, Braszus2020}. Although the exact orientations  of these structures at depth are debated, the surface projections of the Marathon and Mercurius fracture zones and the domain boundary pass through the arc near Guadeloupe and Dominica, close to where the highest seismicity rate is also observed.  
\newline
\noindent
Since 1952, when regional instrumental earthquake recording started, the arc has been covered by internationally operated permanent stations overseen by the University of the West Indies (Seismic Research Center Trinidad): the current \textit{TRN} Network consists of over 50 seismic stations covering the whole arc with further individual stations and local networks (e.g. CU, G, MQ, WI, etc.) from various agencies (e.g. IPGP, KNMI, USGS, etc.). The established extensive island instrumentation helped to significantly improve the regional earthquake data catalog over the years.
\newline
\noindent
Earthquake focal mechanisms (FMs) provide valuable insights into seismic fault structures and regional stress fields. However, no routine operational catalog of focal mechanisms for the region exists. Available catalogs are based on retrospective, time-limited studies. Until 2021, to our knowledge, around 106 focal solutions of various quality were available for the northern part of the arc. Using local island stations, \citep{GONZALEZ2017214} derived 29 solutions using available waveforms of land stations. \cite{Ruiz2013} derived 22 focal mechanisms using first motion polarities from a temporal amphibious ocean-bottom seismic (OBS) experiment in the fore-arc offshore Dominica. The remaining solutions are from various international agencies (e.g., GCMT, Geofone, and USGS), and are constrained by recordings at teleseismic distances.
\newline
\noindent
Due to the shape of the LA island arc, regional land installations are restricted. This inherent coverage may bias the source mechanism solutions using land station data alone.  An offshore extension of the local network is therefore essential to increase the database of regional FMs, particularly for small and intermediate-sized events. Compared to scarce large events with a magnitude greater than 7, more regularly occurring smaller events provide detailed insight into fault structures in the crust and the subducting Atlantic lithosphere. They also offer a chance to study the regional stress field and its variation along and across the arc.
\newline
\noindent
Research projects investigating the physical properties of subduction zones now often incorporate OBS stations (e.g. \citep{Romanowicz1998, Ruiz2013, Cabieces2020, LeonRios2021}). However, the recorded data often provide only a limited source of information for regional moment tensor (RMT) source modeling, for example polarity information in a joined RMT inversion (e.g., \citep{Ruiz2013, Cabieces2020}) or in an inversion restricted to pure double-couple solutions. Two main sources of errors may affect the effective incorporation of OBS data in RMT full-waveform inversions – one are ocean microseisms of which the dominant frequencies around 1.0 - 0.05 Hz \citep{Yang2012} have a strong influence on the horizontal components of OBS stations (supplementary Figure \ref{fig:PSD}). The other source of error is the uncertainty in station locations and alignments of the sensors. Station locations have been assessed using direct arrival times and remaining uncertainties can be accepted if the recorded data are used for studies with wavelengths exceeding them. The influence of alignment, however, is frequency independent and directly affects waveform modeling analyses. Different methods have been developed to estimate the alignment angle of sensors relative to north, including ambient noise correlation \citep{Yang2013}, receiver functions \citep{Janiszewski2015,Lim2017},  P-wave polarity \citep{Wang2016} and its combination with Rayleigh-wave polarization \citep{Doran2017} using natural sources or even artificial ones like ship tracking \citep{Trabattoni2020}. 
\newline
\noindent
Existing RMT inversion routines can handle uncertainties related to the increased noise level of the observations (e.g. \citep{Vackar2017,Pugh2016,Vasyura2020}). However, to our knowledge, none of these published approaches simultaneously consider the alignment uncertainty at an arbitrary station. Here we present a new inversion routine, Am$\Phi$B - Amphibious Bayesian, based on full waveform inversion \citep{Krizova2013} in a Bayesian framework \citep{Duputel2012}. We introduce two covariance matrices – one for data error $C_d$ associated with increased ambient noise and the second for the model error $C_T$ as a function of the sensor alignment uncertainty. The inversion problem (e.g. \citep{Yagi2008, Duputel2012}) is then solved in a uniform tree-sampling search algorithm \citep{Lomax2001}, using the uniform Tape parameterization \citep{TapeandTape2015} for a deviatoric source mechanism.
\newline
\noindent
In this study, we first demonstrate the effects of horizontal alignment uncertainties resulting from both error sources, using synthetic tests of two distinct focal solutions. Following these examinations, we apply our new RMT inversion routine to 45 events in the local earthquake catalog \citep{Bie2019}. We then compile an RMT database that includes newly-derived and pre-existing focal mechanism solutions for the northern part of the Lesser Antilles subduction zone and conduct an inversion for the regional stress orientations.

\section{Method Am$\Phi$B - full waveform RMT inversion routine}
\label{sec:Method}
An earthquake can  be approximated mathematically as a point source by a symmetric second order moment tensor, which consists of six equivalent force couples \citep{Aki2002}. The displacement time-series $d^{obs}_{n,j}(t), j \in [Z,R,T]$ recorded at a surface station $n \in N$ is not only governed by the source mechanism but also the structural and material properties between source (earthquake) and receivers (seismic stations) \citep{Jost1989}. The structural model provides a set of Green’s functions $G_{ij}$ that subsequently construct the synthetic displacement time series $d^{synt}_{n,j}(t)$ in a linear weighted summation of the modeling elements $m_i$ (e.g. moment tensor elements) 
\begin{equation}
d^{synt}_{n,j}(t) = S(t) \cdot \sum_i^I G_{ij} \cdot m_i  + e_n(t)
\label{eqn:nonLinSum}
\end{equation}
with source time function $S(t)$ and residual $e_n(t)$.  Design of $\bf{G}$ and model vector $\bf{m}$ is dependent on the number $i \in I$ of fundamental mechanisms the source is decomposed into (e.g. \citep{ZhaoHelmberg1994,Krizova2013,Dahm2014}). The radial symmetry of a 1D velocity model requires only three pure double couple mechanisms to describe any arbitrary double couple source. Hereby, $\bf{G}$ is designed for displacement traces along pressure, tension and nodal axis, and assembled by incorporating the station azimuth as an additional parameter to the double couple source mechanism defined by strike, dip and rake \citep{ZhaoHelmberg1994}. A full deviatoric model, on the other hand, is more sophisticated \citep{Krizova2013} and uses five fundamental double-couple (DC) mechanisms at the given source-receiver geometry (e.g. \citep{Sokos2008, Sokos2013, Zhu2013}). In the scope of this study we use a combination of both methods for a deviatoric source inversion. Eq.\ref{eqn:nonLinSum} can be written in matrix form
\begin{equation}
d^{synt} = \mathbf{G}\mathbf{m} + E    
\label{eqn:nonLinMat}
\end{equation}
where the source time function $S(t)$ from Eq.\ref{eqn:nonLinSum} is integrated into the Green’s function matrix. The error $E$ is introduced to compensate differences between the synthetics and actual observation and has been used to comprise multiple uncertainties including source specific assumptions such as wrong centroid location \citep{Duputel2012}, variations in source time function \citep{Staehler2014}, fault structure complexities \citep{Yagi2008}, station alignments (this study) or general local background noise (this study). Considering these uncertainties in the moment tensor inversion problem, Eq.\ref{eqn:nonLinMat} can be expressed in an ordinary least squares formulation \citep{Yagi2008}:
\begin{equation}
(\mathbf{G}\mathbf{m}-d^{obs})^T\mathbf{C}_D^{-1}(\mathbf{G}\mathbf{m}-d^{obs}) \rightarrow min.
\label{eqn:OLS}
\end{equation}
To statistically quantify the errors, we introduce a Gaussian distributed probability density. We construct the probability density function (PDF) as stochastic information of the model space:
\begin{equation}
Q(\mathbf{m}) = ke^{-0.5[ (\mathbf{G}\mathbf{m}-d^{obs})^T\mathbf{C}_D^{-1}(\mathbf{G}\mathbf{m}-d^{obs})]}
\label{eqn:Bayesian}
\end{equation}
where $k$ is a normalization factor satisfying $\sum Q(\mathbf{m}) = 1$. 
Here, $\mathbf{C}_D$ represents the covariance matrix of error sum $E$. Likewise, this matrix can be decomposed into the underlying error sources following the error propagation rules
\begin{equation}
\mathbf{C}_D = \mathbf{C}_d + \mathbf{C}_T .
\label{eqn:CD_comp}
\end{equation}
We assume that error $E$ is attributed to the two largest contributions for an OBS deployment: Ambient noise of ocean microseism with data covariance $\mathbf{C}_d$ and station dependent alignment of horizontal components defining model covariance matrix $\mathbf{C}_T$.

\subsection{Ambient Noise}
While land stations might have to deal with strong noise due to human activities inducing frequencies greater than 1 Hz (e.g., \citep{Lecocq2020}), OBS stations are comparably quiet in this range \citep{Yang2012}. Instead, OBS recordings show a much stronger ocean-induced microseism at frequencies between 0.05-1 Hz, which is most pronounced on the horizontal components. The noise at station $n$ is assumed to follow Gaussian distributed white noise $\nu_n(t)$ with mean $\overline{\nu_n(t)} = 0$ and standard deviation $d\nu_n(t)$:
\begin{equation}
\sigma_n = \overline{\nu_n(t)} \pm d\nu_n(t).
\end{equation}
For the construction of the data covariance matrix $C_d$ we follow the design by \cite{Duputel2012}. A simple diagonal matrix design with $C_d = \sigma_n^2 \bf{I}$, $\bf{I}$ being the identity matrix, may result in under-estimations of the uncertainties due to oversampling. This issue is tackled by additionally considering the sampling frequency of the band-pass filtered signal to compensate for this effect:
\begin{equation}
(C_d^{ij})_{n} = \sigma_n^{2}  e^{-|\Delta t^{ij}| / t_{0}} 
\label{eqn:Cd}
\end{equation}
with $t_{0} = 1/f_{min}$ and $\Delta t^{ij}$ being the time difference between sample $i$ and $j$ \citep{Duputel2012}.

\subsection{Station alignment}
\begin{figure}
\centering
\includegraphics[width=1.0\linewidth]{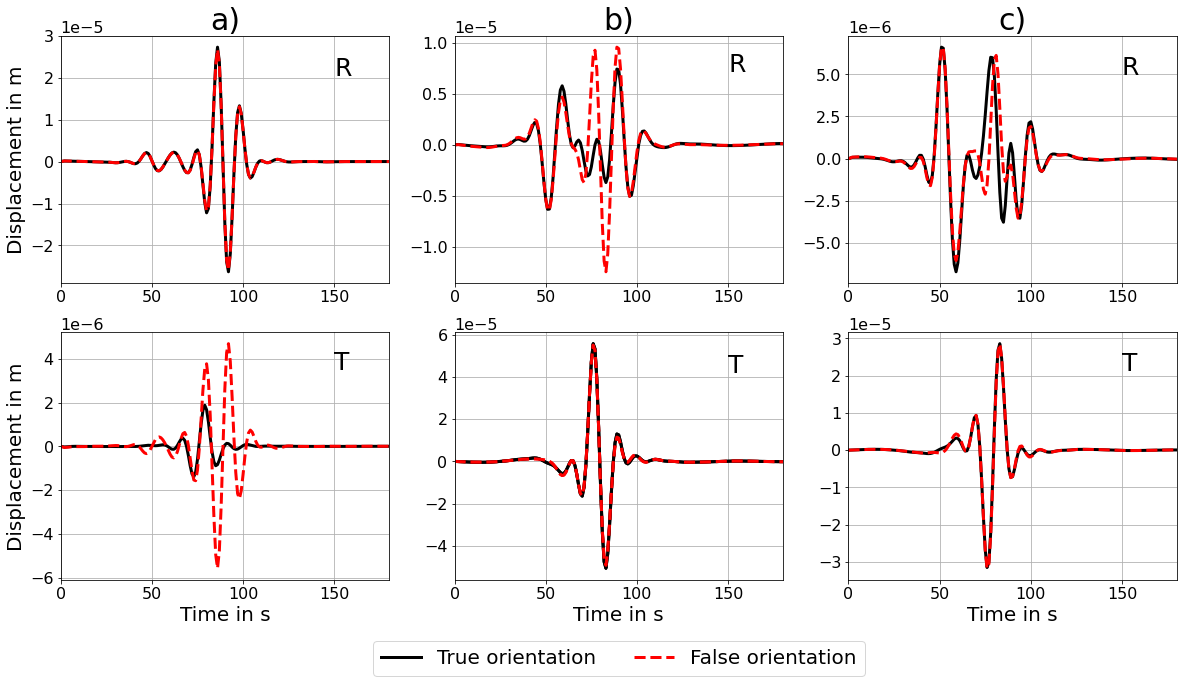}  
\caption{Influence of d$\alpha = 10^\circ$ station alignments of horizontal recordings for different source mechanism and station azimuth $\theta$ from a synthetic test. R = Radial; T = Transverse component.
\textbf{a)} Dip-Slip $(\psi,\delta,\lambda) = (0,90,90)$, $\theta = 80^\circ$ near p-axis ($90^\circ$), energy on T gets minimal 
\textbf{b)} Strike-Slip $(\psi,\delta,\lambda) = (0,90,0)$, $\theta = 20^\circ$ near n-axis ($0^\circ$), energy on R gets minimal
\textbf{c)} Normal Fault $(\psi,\delta,\lambda) = (0,45,-90)$, $\theta = 30^\circ$ near p-axis ($45^\circ$), energy on R gets minimal}
\label{fig:MRot_ex1}
\end{figure}

Aligning a seismometer to true north is an important step during installation to ensure data consistency within a seismic network. While the alignment is usually well-calibrated for land stations, the orientation of OBS is random and has to be estimated after the deployment. Figure \ref{fig:MRot_ex1} displays the influence of a deviating alignment angle of d$\alpha$ = 10$^\circ$ from true north, which we deem a realistic uncertainty in actual networks, for different source mechanisms on the radial (R) and transverse (T) component on stations at different azimuths. 
The three examples show that a false alignment strongly affects at least one of the components depending on station azimuth relative to the pressure, tension, and nodal axis of the mechanism. Correcting the alignment with angle $\alpha_n$ at station $n$ can be expressed by a  rotation of the horizontal components around the vertical axis Z:
\begin{equation}
d_n = \mathbf{R}(\alpha_n) d'_n,
\label{eqn:RotMat}
\end{equation}
$\mathbf{R}$ being the rotational matrix around the vertical axis. 
Eq.\ref{eqn:RotMat} can be interpreted as a linear energy redistribution between the horizontal components R and T scaled by angle $\alpha_n$ at small deviating alignments. In practise, the true alignment angle $\alpha_{n,True}$ cannot be derived exactly but has to be estimated within a reasonable uncertainty range:
\begin{equation}
\alpha_{n,true} = \overline{\alpha}_n \pm d\alpha_n.
\label{eqn:dAngle}
\end{equation}
Uncertainty d$\alpha_n$ can then be translated into a general expression for the covariance matrix for the moment tensor inversion \citep{Duputel2012} by:
\begin{equation}
\mathbf{C}_T = \int  [d(\overline{\alpha}_n) - \overline{d}] [d(\overline{\alpha}_n) - \overline{d}]^T \rho_a da    
\label{eqn:IntRandRot}
\end{equation}
with $\overline{d}$ being the average data vector of a population of random deviating angles according to expression \ref{eqn:dAngle} and $\rho_a$ the Gaussian probability density of the alignment. This equation can be solved by a Monte Carlo approach using a large population of Gaussian distributed uncertainties in the alignment at each station. However, instead of conducting a large-scale random simulation, we benefit from the data gradient following Eq.\ref{eqn:RotMat}, which allows us to consider the sensitivity of d$\alpha$ to the horizontals for a given focal mechanism. Thus, we can give an expression for observations in the vicinity of the true alignment to the north with:
\begin{equation}
d(\alpha_{n,true}) = d(\overline{\alpha_n}) + \nabla d(\overline{\alpha_n})(d\alpha_n - \overline{\alpha_n}).
\label{eqn:dRot_approx}
\end{equation}
Since d$\alpha_n$ is assumed to be generally small, we are able to smooth near $\alpha_n$ and write $\overline{d} = d(\alpha_n)$ in Eq.\ref{eqn:IntRandRot} \citep{Duputel2012}, leading to:
\begin{equation}
C_T = [\nabla d(\overline{\alpha_n})] C_{d\alpha_n} [\nabla d(\overline{\alpha_n})]^T
\label{eqn:CT}
\end{equation}
with $C_{d\alpha_n}$ being the variance of d$\alpha$. Alignment gradient $\nabla d(\overline{\alpha_n})$ is computed numerically as a second order approximation of Eq.\ref{eqn:RotMat} to consider clockwise and anti-clockwise alignment uncertainty:
\begin{equation}
\nabla d(\overline{\alpha_n}) = \frac{d'_n(d\alpha_n) - d'_n(-d\alpha_n)}{2d\alpha_n}.
\label{eqn:derivO2}
\end{equation}

\subsection{uniXtree - uniform X-dimensional tree-importance sampling}
Grid search analysis for seismic source inversion is a straightforward and robust technique to examine complex problems in arbitrary X-dimensional parameter space but at the cost of potentially large computational times. It is therefore important to perform the inversion efficiently by avoiding non-uniformity but at the same time allowing for fast convergence to the global minimum. 
\newline
\noindent
Uniform sampling of the source parameters in an evenly spaced cartesian coordinate system, be it a direct sampling of 6-moment tensor elements or in a parameterization of five source parameters and seismic moment $M_0$, does not result in a uniform distribution of moment tensor solutions \citep{Tape2016}. Likely effects are under-or oversampling of relevant or irrelevant solutions that can cause an increase of the computation time, large sampling-based uncertainties, or even overestimations of local minima.
A possible solution to this issue is the n-D hyperspace method \citep{Tashiro1977} that enables uniform sampling of the moment tensor elements \citep{Staehler2014} on an ellipsoid using 5 independent parameters and the seismic moment. This approach is straightforward as uncertainties become directly apparent on the elements. In further examinations, however, this leads to the general decomposition analysis for the full tensor to study isotropic changes and CLVD (Compensated Linear Vector Dipol) parts as well as the uncertainties in respect to the fault geometry. 
In this study, we employ the approach by \cite{TapeandTape2015} where the source is represented in a geometric framework spanned by five independent angles that are directly connected to the source parameters. Strike, dip and rake are parameterized on a sphere whereas CLVD and isotropic parts are projected onto a lune. In this work, we only focus on deviatoric parameters, hence a general tectonic shear mechanism without volume changes. The remaining space has unique features with different behaviors at the boundaries:
\begin{itemize}
    \item Strike with $\psi \in [0,2\pi]$ 
    \item Dip with $\delta \in [0,\frac{\pi}{2}]$ and in the uniform Tape representation: $h = arccos(\delta)$ with $h \in [0,1]$ 
    \item Rake with $\lambda \in [-\pi,\pi]$
    \item CLVD $\in$ [-100\%,100\%] and in the uniform Tape representation: 
    $u = \frac{1}{3} $arcsin$(3\gamma)$ with $\gamma \in [-\frac{\pi}{6},\frac{\pi}{6}]$
\end{itemize}
Using a direct sampling approach of the Tape parameters, we explore the tree-importance sampling algorithm that is extensively used for different computational problems in e.g. 3D visualizations \citep{Meagher1982,Yamaguchi1984} or in more modern applications together with neural networks \citep{Wang2017_oct}. The application to geophysical problems was first applied by \cite{Lomax2001} for the octant sampling of the 3D hypocenter location. Compared to generally used techniques like the Monte Carlo or Metropolis-algorithm, it has been shown that an oct-tree algorithm is more global and complete while only depending on a few parameters. The general tree data sampling structure is recursively subdividing X-dimensional spaces into $2^X$ sub-volumes. For 1-dimension the tree structure results in bisecting the search interval, whereas higher dimensions like in 3D lead to 8 subvolumes called octants (hence commonly referred to as an oct-tree). It drastically reduces the number of samples by an order of $10^5$ with respective computational time while allowing for examinations of the whole model space in a robust manner. For the application to moment tensor inversion, we sample in a 4D space spanned by $\psi$, $h$, $\lambda$, and $\gamma$. Probability $P_i$ of a source solution is then defined by the PDF $Q(\mathbf{m})$ (Eq. \ref{eqn:Bayesian}) and weighted by the represented sub-volume $V_i$ with:
\begin{equation}
P_i = V_i \cdot Q(\mathbf{m})   
\label{eqn:Probability}
\end{equation}
The uniXtree approach deployed, enables an efficient way to increase resolution within the full model space around highly probable solutions while only sampling coarsely volumes of less importance.

\section{Data}
\label{sec:DataModel}
\subsection{VoiLA OBS deployment}
\label{subsec:VoiLA}
Between March 2016 to May 2017, a regional OBS network was installed as part of the NERC-funded international multidisciplinary consortium project - VoiLA, to study the \textbf{Vo}latiles \textbf{i}n the \textbf{L}esser \textbf{A}ntilles Island Arc \citep{Goes2019}. The network consists of 34 broad-band sensors deployed in the fore- and the back-arc area \citep{Collier2015} along the arc with a focus on the northern parts (Fig. \ref{fig:Map1}). During the operation time, 381 seismic events were located with high accuracy and assigned with local magnitudes \citep{Bie2019}. The stations were located at an average water depth of $\sim$2800 m with individual instruments as shallow as 812 m and as deep as 5054 m. Applying the empirical relation between water depth and the peak-frequency of related ocean induced microseismicity (Fig. \ref{fig:PSD}) \citep{Yang2012}, we find dominant ocean microseism signals between 0.008 Hz to 0.01 Hz. Considering the resolution restriction of our local 1D velocity model \citep{Bie2019}, the usable frequency range is set between $f_{min} = 0.03$ Hz and $f_{max} = 0.1$ Hz. Approximate horizontal sensor orientations were determined using the Rayleigh-wave polarisation analysis method of\citep{Doran2017}; however, formal uncertainties in these azimuths, d$\alpha_n$, can reach up to $12^\circ$. 

\subsection{Data selection}
In this study, we focus on the northern part of the LA arc. The target area spans between $14.4^\circ$N to $18.5^\circ$N and between $60.0^\circ$W in the fore-arc to $62.75^\circ$W in the back-arc region (Fig. \ref{fig:Map1}; black box). For later interpretive approaches, the area is further split into upper plate crustal (z < 33 km), slab interface (33<z<100km) and deep intraslab (z > 100 km) layers as well as a northern (> $\sim16.5^\circ$N) and southern sector. Initial event selection is based on local magnitudes and reported source locations by \cite{Bie2019}. For deep hypocenter locations, a minimum magnitude of ML4.1 has shown to have a sufficiently high signal-to-noise ratio (SNR) while for crustal events (up to 33 km) we were able to examine events down to ML3.8. The minimum number of stations used in the inversions is set at 5, but the vast majority of the events have only a sufficiently good SNR on vertical components. 45 out of the 381 cataloged events \citep{Bie2019} fulfill these conditions and display a sufficient enough SNR on at least the vertical components. The six largest events feature a focal solution in the USGS database (Tab.\ref{tab:RefEvents}) and partially the GCMT catalog which we use for calibration. For the modeling of these events, we use the reviewed source information by \citep{Bie2019} and for the construction of the Green Functions matrix $\mathbf{G}$, the therein published local 10 layered 1D velocity model. In total 39 events are new moment tensor solutions not already published in any database. We define two sets of events - the first comprising six events with FMs reported by USGS and GCMT belong to a reference event set, and the rest assigned to a new event set.

\section{Synthetic tests}
\label{sec:Synthetic_examination}
To evaluate the robustness and accuracy of our inversion algorithm, we conduct a series of synthetic tests. In the first part, we study a synthetic strike-slip mechanism recorded within the station arrangement of the VoiLA OBS network, to compare the standard linear source inversion with our new Bayesian approach that takes into account the station alignment and ambient noise individually. In the second part, we examine how our inversion routine performs with various source mechanisms and the number of OBS recordings. 

\subsection{Influence of station alignment}
\begin{figure}
\centering
\includegraphics[width=1.0\linewidth]{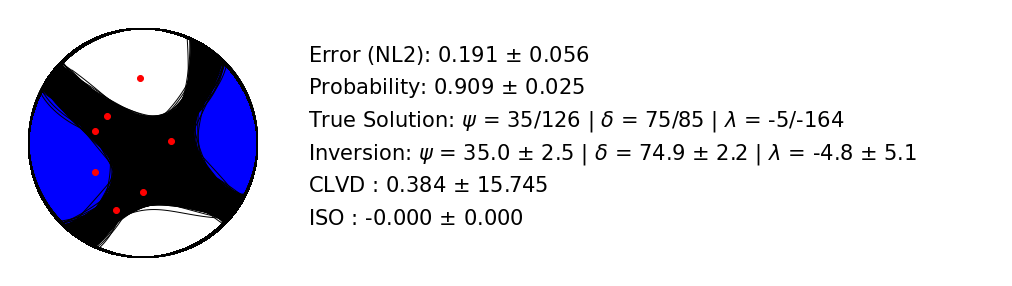}
\caption{Influence of station alignments on a distinct strike-slip source solutions. Effect of 10000 Gaussian distributed station alignments with mean $\overline{\alpha}_n = 0$ and standard deviation $d\alpha_n = 15^\circ$ on a normal fault and strike-slip mechanism.}
\label{fig:MRotTest_SS}
\end{figure}

Let us assume a well-distributed sub-network within the given VoiLA constellation consisting of 7 receivers. The source is located at its center and set as a pure strike-slip with ($\psi,\delta,\lambda$) = (35,75,-5) (Fig. \ref{fig:MRotTest_SS}). We simulate 10,000 realizations with a normal distributed alignment error at $\overline{\alpha}_n = 0^\circ$ and $d\alpha_n = 15^\circ$ for each station and invert for the source mechanism in a linear L2 inversion. As expected, we observe variations between true and inverted mechanisms in strike, dip, and rake whereas the mean is at the true model. A noticeable change can be observed in the appearance of a non-neglectable CLVD component with a mean near zero and uncertainty of 15\%. A CLVD acts as an amplitude compensation on the waveforms to adjust for missing or excessive mean energy content that cannot be fully described by a pure double-couple model. In the case of false station alignments, the amplitude discrepancies on the horizontals between synthetics and observables are governed by the energy redistribution following Eq.\ref{eqn:RotMat}. While the mean CLVD in a large population of solutions is still around zero, the increased (positive or negative) uncertainty may be mistakenly interpreted as an indicator of erroneous assumptions in the wave-speed model or a more complex source structure. Those assumptions might lead to a complex re-evaluation of the source and occur as a systematic error for projects using the network. A simple linear inversion algorithm is thus not able to handle such an error and will return large uncertainties. 

\subsection{The influence of ocean noise}
\begin{figure}
    \centering
     \begin{subfigure}[b]{0.48\textwidth}
         \centering
         \includegraphics[width=\textwidth]{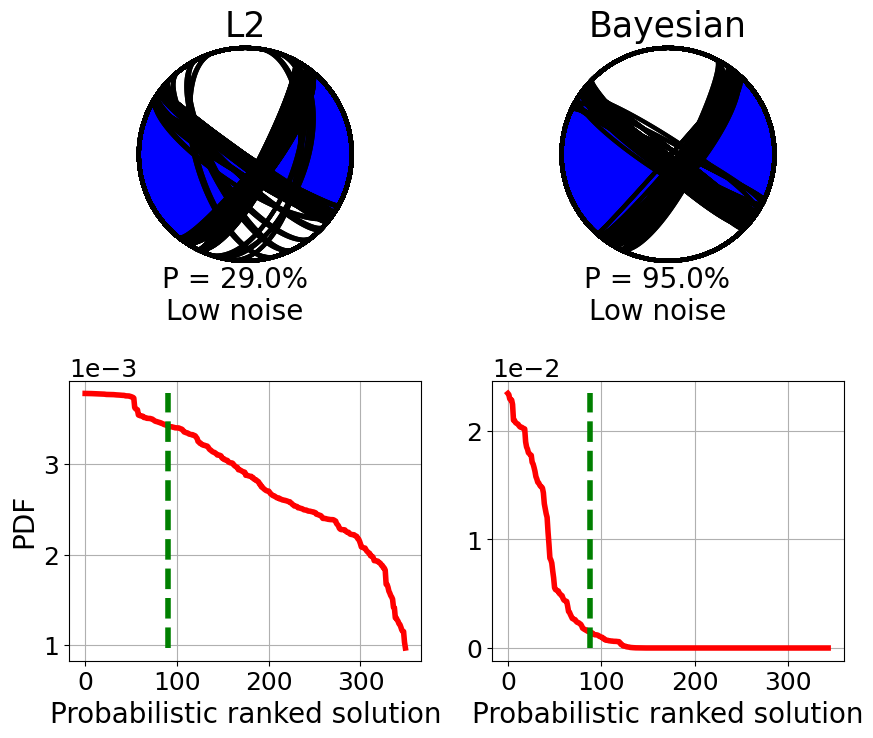}
         \caption{Low noise}
         \label{fig:Cd_a}
     \end{subfigure}
     \hfill
     \begin{subfigure}[b]{0.48\textwidth}
         \centering
         \includegraphics[width=\textwidth]{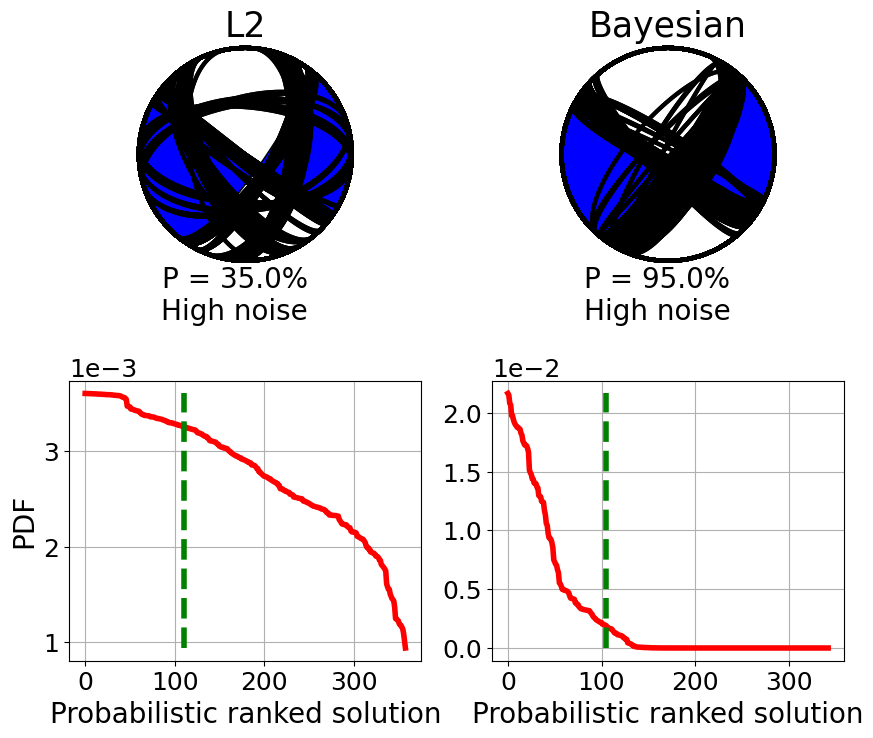}
         \caption{High noise}
         \label{fig:Cd_b}
     \end{subfigure}
\caption{Pure DC inversion results for a strike-slip mechanism using a L2 and a Bayesian error formulation. Simulations are performed for low noise a) and a high noise level b). True source solution is displayed in blue with top solutions marked by black lines. Below each beachball is the sorted probability density function, the green vertical dotted line marks the top 95\% of the Bayesian solution and the corresponding number of samples in the L2 inversion at the same noise level.
}
\label{fig:syntTest2}
\end{figure}

In extreme cases, noise can lead to unfavorable destructive or constructive alterations of the waveforms leading to misinterpretations of local minima with subsequent ambiguous solutions and large uncertainties. We examine the difference between our Bayesian error formulation and an L2 (no covariance matrix) cost function (Eq. \ref{eqn:Bayesian}). 
Figure \ref{fig:syntTest2} displays the results for a Bayesian (data covariance matrix $C_d$) formulation and a L2 cost function at two different signal-to-noise (SNR) levels. The top row shows the FMs for the true mechanism (blue) and the best inversion results (black). The number of the best solutions depends on the 95\% confidence interval of the Bayesian run and is marked as a green dotted line in the PDF graphs. Importance sampling using a Bayesian formulation suppresses improbable solutions and highlights one with high probabilities as shown by the narrow PDF. This suppression is not applicable for an L2 formulation which results in a much wider PDF covering a larger volume in the parameter space representing potentially different solutions.
\newline 
\noindent 
For low (Fig. \ref{fig:Cd_a}) and high (Fig. \ref{fig:Cd_b}) noise the Bayesian formulation yields the true source mechanism with a corresponding level of uncertainty.
In the case of L2, however, we observe also a possible thrust fault solution parallel to the tension axis of the true strike-slip mechanism. Tensional and compressional areas of both mechanisms show similarities in the recorded waveforms and can lead to wrong identification for unfavorable network configurations. If we increase the noise level, the L2 approach returns two probable solutions in the form of a normal and thrust fault parallel to the tension and pressure axis, respectively. 

\subsection{Influence of the VoiLA network geometry}
Based on the network geometry of the VoiLA network a subset consisting of a given number of stations within a source-receiver distance of 350 km are randomly chosen to perform source inversion for an arbitrary double-couple mechanism. We choose ev123 - an earthquake located in the middle of the network (Tab.\ref{tab:RefEvents}). To quantify the uncertainty of the fault geometry, we use the Kagan angle \citep{Kagan1991} as well as percentile changes in the CLVD. Measured discrepancies between the true source mechanism and the inversion results are displayed in a normalized ascending sorted misfit curve. To compare the performance of different inversion approaches, we calculate a reduction percentage between the L2 and the Bayesian formulation:
\begin{equation}
   reduction = 100 \cdot \left(1- \frac{\sum pdf_{Bayesian}}{\sum pdf_{L2}}\right)
\end{equation}

\subsubsection{Performance using data covariance $C_d$}
The Bayesian error formulation employs only the data covariance matrix $C_d$. In the construction of the matrix, 50\% of the simulated vertical peak noise amplitude of each trace is set as $\sigma_n$ in Eq.\ref{eqn:Cd}. The station-wise simulated noise on the horizontal components is then re-scaled with a constant factor between the horizontal to the vertical peak noise amplitude (H2V). Our tests were conducted for 4, 6, and 8 random distributed stations but with a fixed source location. To restrict the number of free parameters, we keep the magnitude fixed. 
\newline 
\noindent 
Figure \ref{fig:Noise_Test} summarizes the results of the random sample study at all setups. The average discrepancy in the Kagan angle and CLVD percentage increases proportionally to the H2V factor and with decreasing number of stations. Significant differences appear between the L2 and the Bayesian formulation, where the Kagan angle and CLVD percentage is predominantly smaller using the latter approach. Especially the error to the true CLVD could be reduced (relative reduction of the areas below the curves) by around 48\% at a horizontal H2V factor of 1.0 and up to 59\% at larger noise with H2V factor 2.0. Reduction of the Kagan discrepancy is again smaller for H2V factor 1.0 at 20\% to 28\% but much better at H2V factor 2.0 with a reduction of up to 41\%. In all settings, however, we can observe significant outliers in the Bayesian approach of the Kagan angle at H2V factor 1.0. This suggests a limitation of a Bayesian formulation solely based on the data covariance \citep{Duputel2012} and is presumably due to a number of low noise realizations. 

\subsubsection{Performance using data and model covariance $C_D$}
\begin{figure}
\centering
\includegraphics[width=1.0\linewidth]{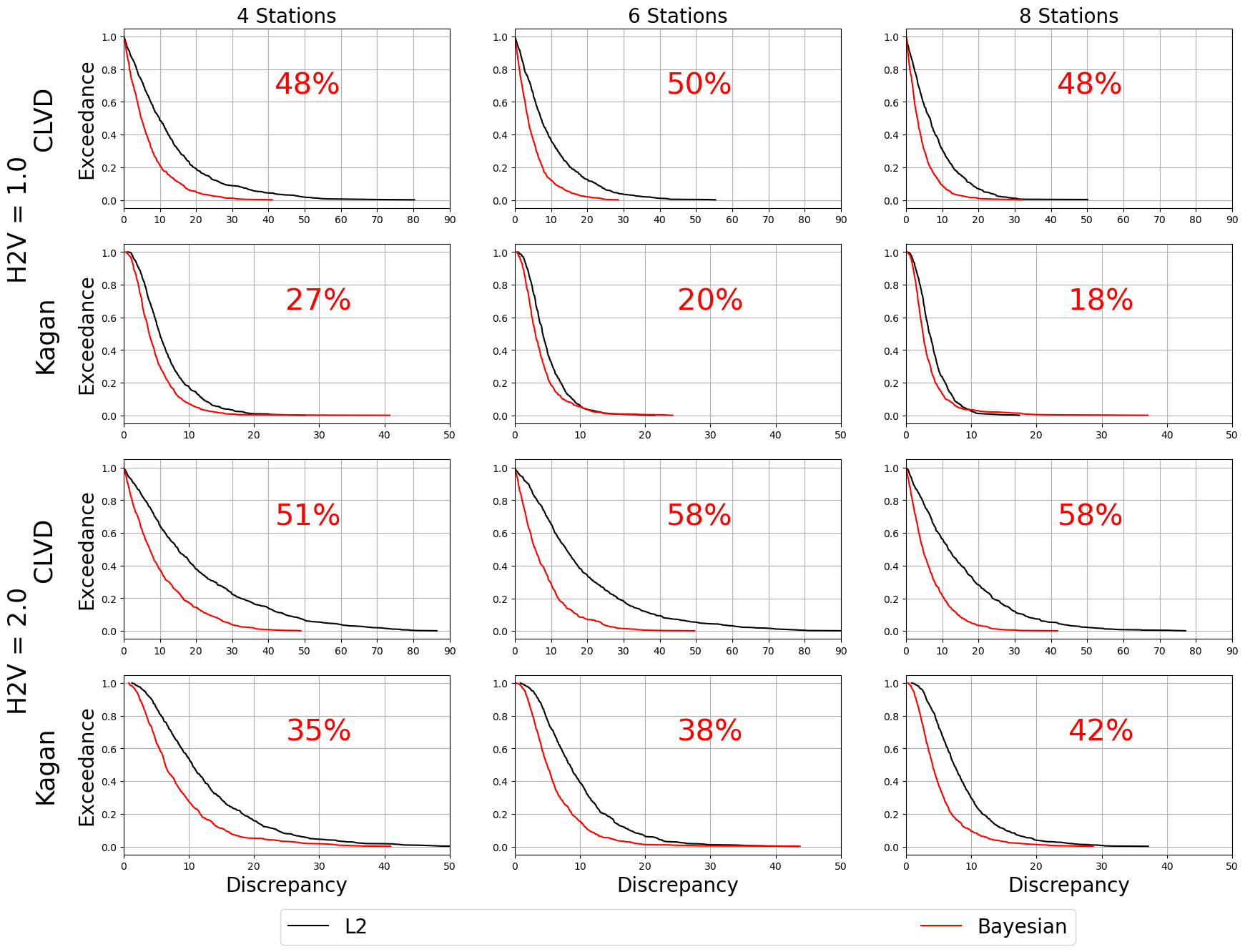}
\caption{Synthetic noise test with base amplitude level at 50\% of vertical peak signal. Surveyed in a random study (station location, noise, and true source mechanism) are the discrepancies in Kagan angle and CLVD percentage between the true solution and the inversion results. The examinations were conducted for a random network consisting of 4, 6, and 8 stations at a horizontal H2V amplification factor of 1.0 and 2.0. Displayed PDF curves represent the ascending discrepancy normalized at the absolute number of the result samples (>1000). Marked as black curves are the results for a simple linear inversion while red is a Bayesian formulation using a data covariance matrix $C_d$ in a deviatoric tree-importance sampling. Corresponding percentage values are the relative reduction of the curve integral between the L2 to the Bayesian formulation.}
\label{fig:Noise_Test}
\end{figure}

\begin{figure}
\centering
\includegraphics[width=1.0\linewidth]{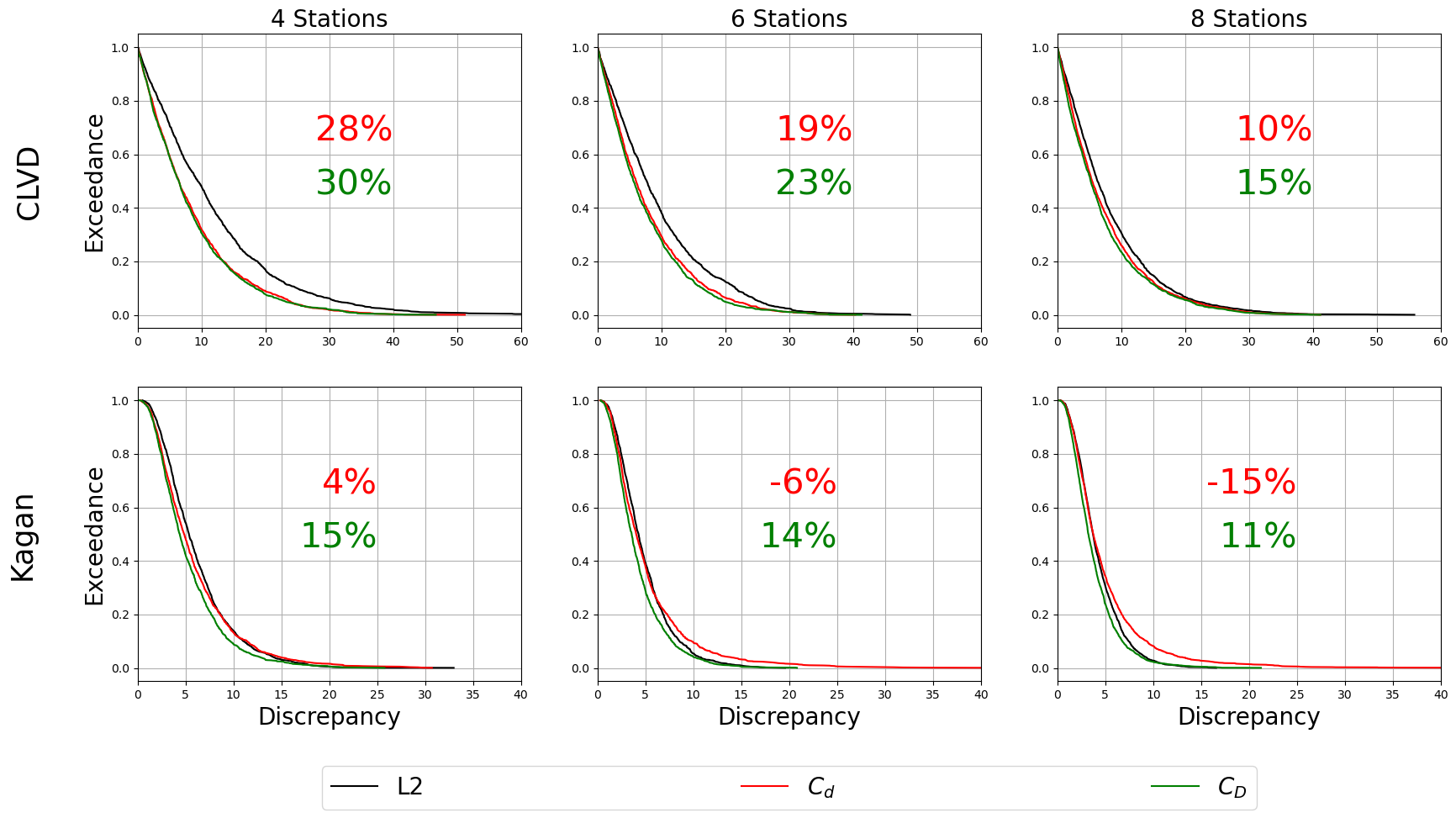}
\caption{Synthetic alignment test at 30\% vertical base amplitude with horizontal to the vertical peak noise amplitude (H2V) factor of 1.0 and $10^\circ$ mean Gaussian distributed station alignment angle. Displayed PDF curves follow the same notation as in figure \ref{fig:Noise_Test} but with a fourth curve (green) displaying the results of a Bayesian $C_D$ formulation.}
\label{fig:MRot_Test}
\end{figure}

Station alignment on the observed data affects the full waveform information on the horizontal components, hence noise and transient signal. 
The model covariance matrix $C_T$ is constructed based on the synthetics, which display the pure transient signal without any noise. This allows us to directly examine the uncertainties on the earthquake signal itself, while noise is handled by the data covariance matrix $C_d$ (Eq. \ref{eqn:CD_comp}). For the construction of $C_T$, we assume an average alignment angle d$\alpha_n$ (Eq. \ref{eqn:CT}) of $10^\circ$ at all stations within a chosen network configuration. As this magnitude of d$\alpha_n$ represents only a small variation, the inversion problem can be treated in a first-order approximation as a linear energy redistribution between the horizontal traces (Eq. \ref{eqn:dRot_approx}). 
\newline 
\noindent 
Given that alignment mainly affects the horizontal components with low energy content (Fig. \ref{fig:MRot_ex1}), we need to consider a low noise level. This assumption stands valid, as small magnitude events within a comparatively high horizontal noise level, as is the case for OBS recordings, are mostly restricted to vertical recordings and hence are independent of this error source. In contrast to that, large events with a good signal-to-noise ratio are more affected by small alignment uncertainties as there are potential non-negligible inferences between the two error sources. 
\newline 
\noindent 
We design a synthetic study for a base noise of 30\% of the vertical trace and no additional amplification H2V factor on the horizontal components (H2V factor 1.0). General settings are adopted from the previous noise examination for 4, 6, and 8 random distributed stations.
\newline 
\noindent 
Figure \ref{fig:MRot_Test} summarizes the results of the random study at all surveyed setups. Displayed curves follow the same notations as in figure \ref{fig:Noise_Test} but with a third one representing the discrepancy of a $C_D$ sampling. For the CLVD we observe a much smaller reduction between the L2 and Bayesian inversion. While the noise conditions are the same in the previous tests with horizontal H2V factor 1.0, it becomes apparent that changes are attributed to the additional station alignment. Differences, however, can be observed between a $C_d$ inversion and and a $C_D$ inversion (\ref{eqn:CD_comp}). While comparatively small, the $C_D$ setting is generally better compared to the $C_d$ inversion and increased with the number of stations. For the Kagan angle, we observe a strong worsening of the $C_d$ inversion in contrast to a $C_D$ inversion. Like in the previous test (Fig. \ref{fig:Noise_Test}) low probable outliers become noticeable in a $C_d$ inversion but do not appear in the $C_D$ inversion.

\subsubsection{Effects of restriction on vertical recordings}
\begin{figure}
\centering
\includegraphics[width=1.0\linewidth]{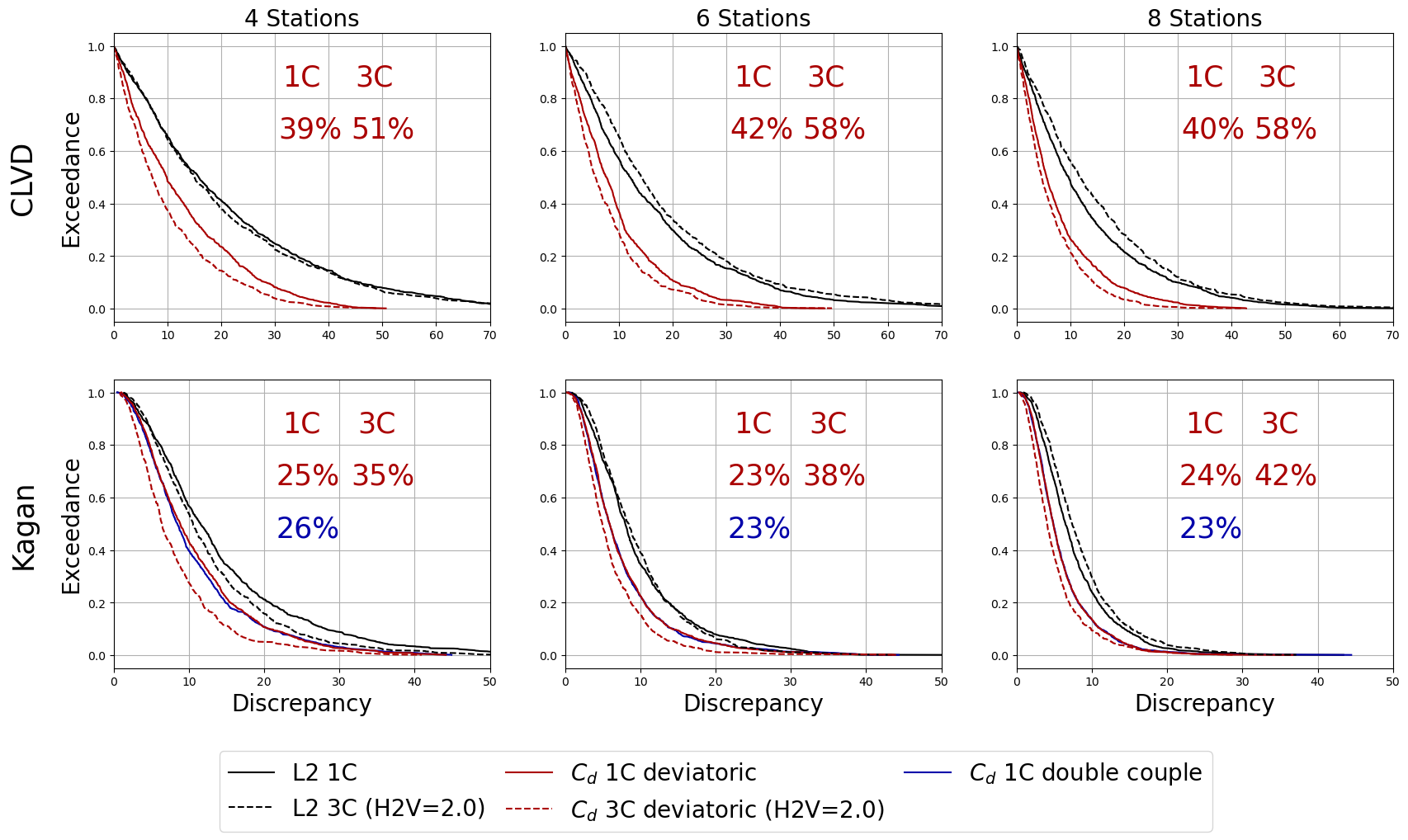}
\caption{Comparison of a vertical trace (1C) data restricted inversion versus a three component (3C) inversion in a synthetic noise test at 50\% vertical base amplitude. Displayed PDF curves follow the same notation as in figure \ref{fig:Noise_Test}.}
\label{fig:Vert_Test}
\end{figure}

For most of the earthquakes in our OBS database, we expect a limitation to the vertical recordings, especially for events at the lower end of the resolvable magnitude range (Mw < 4.8). In a final synthetic study, we examine the effects of a vertical component analysis in comparison to a three-component analysis. In practice, most inversions of small magnitude events are not only limited on the vertical recordings but also to a comparable small and potentially badly distributed network. A full deviatoric description of such sources might hereby be misleading as the CLVD affects the amplitudes of all three components on a different scale. We hereby follow our previous setup with a base noise level of 50\% of the vertical peak amplitude. 
\newline 
\noindent 
As before we can observe a significant reduction of the CLVD discrepancy (Fig. \ref{fig:Vert_Test}), but 10\% to 20\% smaller compared to an inversion employing three components (Fig. \ref{fig:Noise_Test}). In an absolute comparison between a vertical and three-component inversion, we notice a drastic worsening compared to the case with an H2V factor of 1.0 and a similar level at H2V factor 2.0. Assuming that a restriction on only vertical data is attributed to an H2V factor exceeding 2.0, inversion results can be expected to be generally better compared to using all three components. A similar pattern can be observed for the Kagan angle. Interestingly the reduction of the discrepancy in the Kagan angle using a pure double couple inversion becomes similar to that of a deviatoric inversion. 

\section{Results}
\label{sec:Results}
\subsection{Reference event set}
During the VOILA deployment, various international agencies (e.g., USGS, GCMT, GEOPHONE) reported source mechanisms for 6 local earthquakes. These six events define our reference event set, which is used to test the performance of our inversion algorithm with real data. We find large discrepancies in the source depth between the USGS locations and the relocated hypocenters of \cite{Bie2019}, that we fixed for all inversions. All derived MT solutions show a high degree of similarity with published solutions. In the following, we will highlight the inversion results and detailed comparison for the three largest events (Tab.\ref{tab:RefEvents}).
\newline 
\noindent 
Event ev143, the largest earthquake in our observation period with a moment magnitude of Mw5.9, occurred on February 03, 2017, at 19:54:22.86 UTC at a depth of around 50 km within the fore-arc region east of northern  Martinique. We use a total of 19 OBS stations for the inversion but are restricted to using mainly vertical components as most horizontal components exhibit saturation due to strong ground motion in the near-field.
The derived mechanism depicts a thrust fault with (strike,dip,rake) = (341,55,87) and a CLVD of -2\%. Our result slightly differs in the dipping angle (by about 11 degrees) compared to the USGS solution (334,66,84) but shows much higher similarity to the GCMT solution (334,59,86). 
\newline 
\noindent 
Event ev123 occurred on October 18, 2016, at 22:08:13.75 UTC offshore Martinique. With a depth of approximately 160 km and a moment magnitude of Mw5.6, it is the strongest earthquake in this region since the 2007 Mw7.4 Martinique event. Its central location provides good azimuthal coverage and good SNR on all three components of at least 7 stations. We utilized a $C_D$ Bayesian inversion, as we have a sufficient SNR on all three components. Our results (246,42,-93) indicate a trench perpendicular normal fault that is in good agreement with the solutions obtained by USGS (251,43,-100) and GCMT (248,44,-106) but with a slightly increased CLVD of 9\% in our solution. 
\newline 
\noindent 
The third event, ev152 with a magnitude of Mw5.0, occurred on April 17, 2017, at  06:25:10.47 UTC at a depth of 21 km. It is part of the most active cluster in the northern shallow fore-arc offshore Antigua during our deployment. The derived mechanism displays a trench parallel normal fault with a small positive CLVD part of 2\% but shows discrepancies in the double-couple part between our (133,37,-87), the USGS (140,68,-70), and the GCMT (106,40,-125) solutions. The remaining three events are only listed in the USGS database but are generally consistent with our solutions (Tab.\ref{tab:RefEvents}).


\subsection{Combined MT catalog for northern LA subduction zone}
\begin{figure}
\centering
\includegraphics[width=1.0\linewidth]{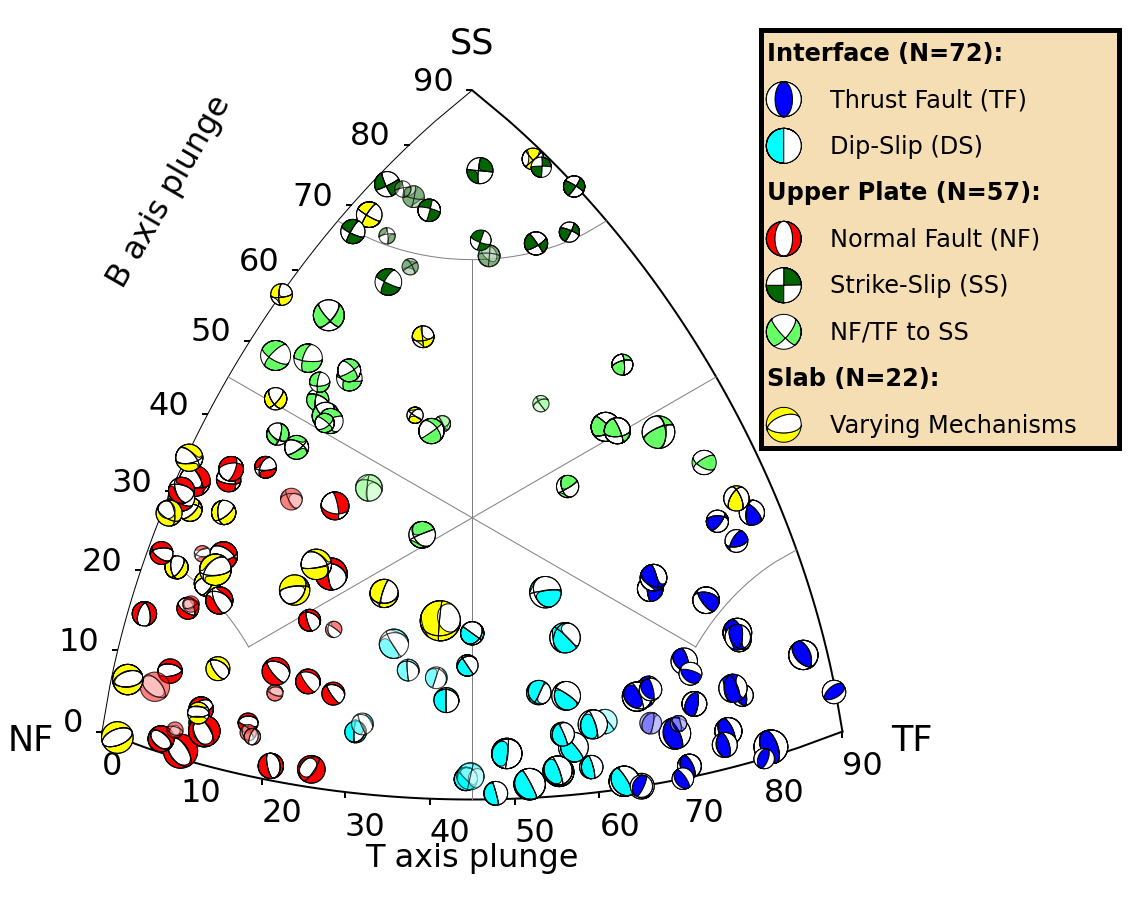}
\caption{Focal Mechanism Classification (FMC) diagram. Displayed are all available focal solutions in the target area. The mechanisms are split into 6 groups manually, based on their locations in the FMC diagram: normal faults (NF; red), thrust faults (TF; blue) dip-slips (cyan), strike-slips (SS; dark green) transition events to strike-slips (NF/TF to SS; light green) and a special group containing all events below 70 km (yellow). Events without certain depth information (e.g. \citep{GONZALEZ2017214}) are displayed in faded colors. The diagram was created using a modified python routine based on \cite{Alvarez2019}.}
\label{fig:FMC}
\end{figure}

\begin{figure}
\centering
\includegraphics[width=1.0\linewidth]{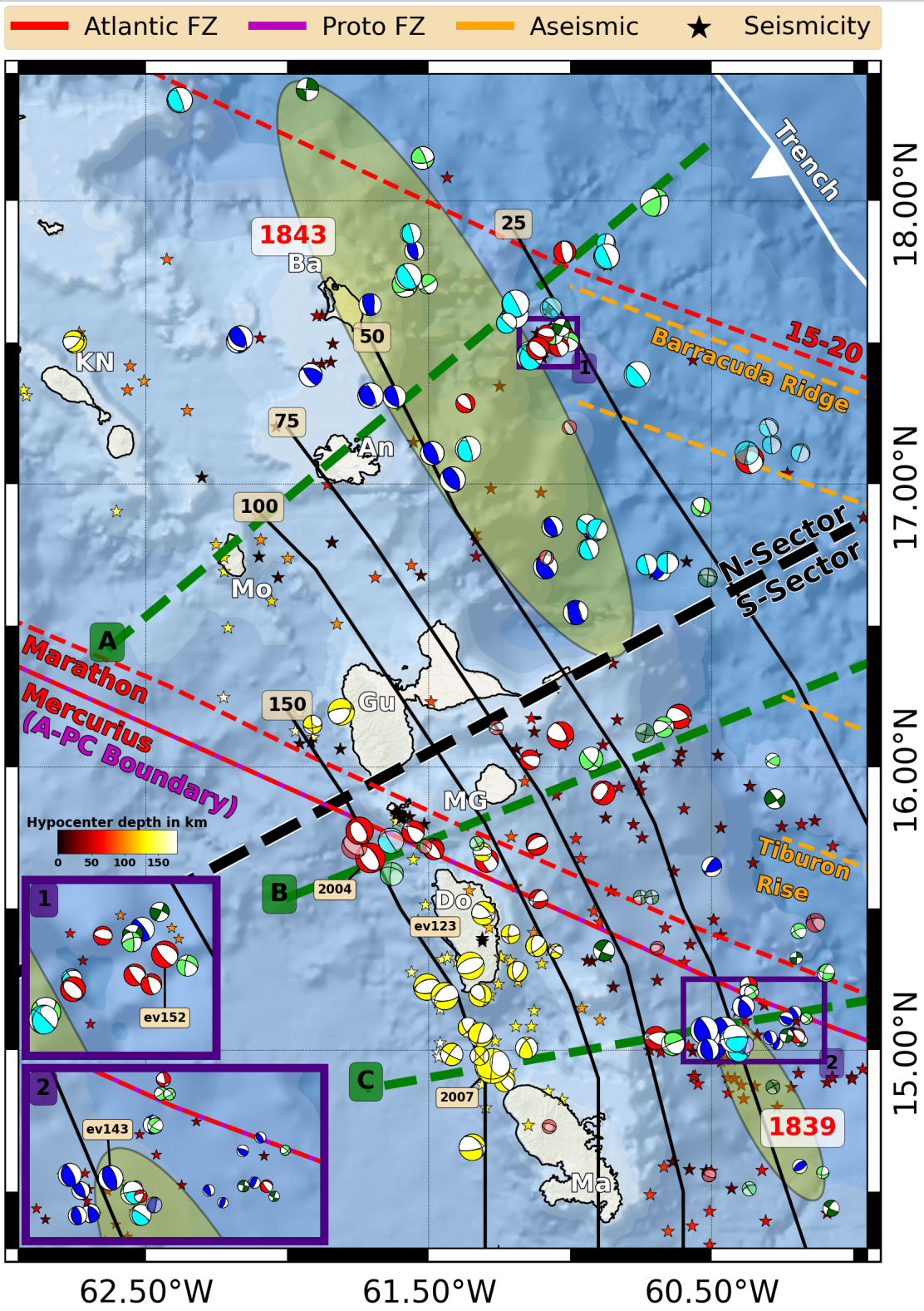}
\caption{Distribution of all existing focal mechanisms from USGS, GCMT, \cite{GONZALEZ2017214} and this study corresponding to the color-coding displayed in the FMC-diagram (Fig. \ref{fig:FMC}). The map covers the islands of Antigua (\textbf{An}), Barbuda (\textbf{Ba}), Saint Kitts and Nevis (\textbf{KN}), Monserrat (\textbf{Mo}), Guadalupe (\textbf{GU}), Marie-Gelante (\textbf{MG}), Dominica (\textbf{Do}) and Martinique (\textbf{Ma}). Depth profiles (green), subducted fracture zones (red: Atlantic; magenta: Proto-Caribbean), oceanic ridges (orange) and trench (white) are introduced in Fig. \ref{fig:Map1}. Displayed ellipsoidal patches represent the locations of the 1839 and 1843 events \citep{GONZALEZ2017214}. Black lines represent the depth profiles of the slab model and depth colored stars the re-located seismic events recorded during the VoiLA deployment \citep{Bie2019}.}
\label{fig:ResMap}
\end{figure}

We compile a catalog of 151 MTs that includes 29 local solutions by \citep{GONZALEZ2017214}, 22 from \citep{Ruiz2013} who focused on a small area east of Dominica and Martinique, 55 teleseismic solutions from various agencies (GCMT, Geofone, and USGS), and 45 derived by Bayesian inversion of OBS waveforms in this study. We mapped this catalog into a focal mechanism classification (FMC) diagram \citep{Alvarez2019} and the crustal events are divided manually into 5 groups based on their FMC location and mechanisms (Fig. \ref{fig:FMC}). All events below 100 km depth form the sixth group that contains a variety of mechanisms but is dominated by NFs. We plot all events (color-coded for each group) in map view (Fig. \ref{fig:ResMap}) together with key features such as projected fracture zones, slab depth-contours, and projected rupture areas of the 1839 and 1843 events. We show three cross-sections (A,B,C) that intersect important tectonic/geodynamic features. The unprecedented quantity of MTs enables a detailed analysis of the earthquake source mechanisms within our target area.
\newline 
\noindent 
\textbf{Plate interface seismicity}\\ 
Seismic activity at the plate interface is marked by thrust events (dark blue mechanisms) at $50 \pm 10$ km depth and shallow dipping thrust events (light blue mechanisms) close to the trench (Fig. \ref{fig:ResMap}). The dip angles follow the subducting slab geometry. In the northern area ($16.5^\circ$N - $18.5^\circ$N) particularly, the thrust events at various depths define a wide seismogenic interface between the subducting and overriding plates. This zone overlaps with the estimated rupture area of the 1843 M8.5 earthquake - the largest megathrust event in the LA recorded in history \citep{Feuillet2011b}. To its south ($15.0^\circ$N - $16.5^\circ$N), a lack of thrusting and shallow dipping events is consistent with previous studies [e.g., \citep{GONZALEZ2017214}]. To the east of Martinique ($15.0^\circ$N,$-60.5^\circ$W), several thrust events are found within the estimated rupture area of the 1839 M7.5 megathrust event \citep{Feuillet2011b} (Fig. \ref{fig:ResMap}, zoom box 2), and there are no shallow dipping thrust events near the trench as found in the north.
\newline 
\noindent 
\textbf{Upper-plate seismicity}\\
Normal faulting (red mechanism) dominates the upper-plate seismicity, with sporadically scattered strike-slip (dark green mechanism) and oblique normal/thrust faulting to strike-slip (NF/TF to SS) events (light green mechanism). The northernmost sector of our target area shows an larger number of oblique events relative to the south. While the occurrence of oblique events decreases towards the south, normal faulting events become more frequent, especially near event ev152. The highly active cluster produces at least five ML > 3.9 events in a day \citep{Bie2019} and features a mixture of normal and NF/TF to SS transitional events (Fig. \ref{fig:ResMap}, zoom box 1). With an average depth of 20 km, this cluster is near the plate interface where we can also observe several shallow dipping thrust events just below and surrounding it. Normal faulting events in the crust increase in the southern sector particularly between Guadeloupe and Dominica ($15.8^\circ$N,$61.65^\circ$W). A prominent event of this cluster is the shallow 2004 Mw6.3 Les Saintes earthquake which ruptured part of an arc-parallel en echelon fault system \citep{Feuillet2011a}. The mainshock triggered up to 30,000 aftershocks within the following two years and is one of the largest instrumentally recorded seismic events in the Lesser Antilles \citep{Feuillet2011a}. We observe additional normal faulting events close to the interpreted upper plate Moho (Fig. \ref{fig:XSection}, profile B), striking in the trench perpendicular direction. Further south, large-magnitude crustal events are rare.
\newline 
\noindent 
\textbf{Slab seismicity}\\
The deep group covers all seismic events with depths greater than 100 km and these locate within the subducted slab. In the northern sector, only three deep events were observed which may be attributed to comparably small magnitudes and unfavorable station coverage in these parts of the arc. The most active zone is between Dominica and Martinique. At $\sim$150 km depth, this area hosts the majority of strong (Mw>4.0) deep earthquakes along the Lesser Antilles Arc within the instrumented period. MT solutions show highly complex faulting structures consisting mostly of normal faults but also a few strike-slip faults and one thrust fault. The normal faults display different fault orientations but are predominately aligned perpendicular to the trench. The largest earthquake within this group is the 2007 Mw7.4 Martinique earthquake, the to-date strongest instrumentally recorded seismic event in the Lesser Antilles. The large $\sim$95\% CLVD part of this event hints at a much more complex source mechanism than pure normal faulting.

\subsection{Stress inversion}
The state of stress in the Lesser Antilles is governed to first order by the subduction of the Atlantic oceanic lithosphere beneath the Caribbean Plate. In most cases, especially for small and intermediate-sized events, rupture occurs on pre-existing weak faults that do not necessarily coincide with the current main stress axis \citep{Plenefisch1997}. The origin of small-scale stress perturbations includes anisotropy, fluid pressure variation, or mineral heterogeneity \citep{Plenefisch1996}. Large events in the subducted lithosphere are likely related to the reactivation of pre-existing faults that formed in past stress conditions (e.g. \citep{Delescluse2008,Garth2014}). To gain insights into the local stress conditions, we perform stress inversions for the interface, crustal, and deep event groups based on our combined catalog.
We use the STRESSINVERSE package \citep{Vavryuk2014} which is based on the approach by \cite{Michael1984} but in an iterative joint inversion, reducing the ambiguity problem due to the identification of active and auxiliary planes. STRESSINVERSE takes as input the source geometry (strike/dip/rake) of a group of events and returns the principal stress directions with $\sigma_1 > \sigma_2 > \sigma_3$ and the scalar quantity $R$  ($0 \le R \le 1$) describing the magnitude of $\sigma_2$ relative to the others \citep{Gephart1984}:
\begin{equation}
    R = \frac{\sigma_1 - \sigma_2}{\sigma_1 - \sigma_3}.
\end{equation}
We assume for each inversion run a mean deviation of $10^\circ$ in a realization of 250 normally distributed variations.
\newline 
\noindent 
Initial stress inversions were performed for the plate interface events and upper plate events separately. The former group includes 50 thrust MTs with the inversion showing a compressional stress regime with $R = 0.52$. The direction of $\sigma_1$ is trending $61^\circ$ and plunging $25^\circ$ (Tab.\ref{tab:StressInv}) in good agreement with the local slab model \citep{Bie2019} which depicts an average dip of $30^\circ$ and a mean azimuth of about $60^\circ$.
In the upper plate , inversion of 73 MT solutions depicts a trench parallel extension stress regime with $\sigma_3$ azimuth at $183^\circ$ and plunge $3^\circ$ and  $R = 0.87$. The southern sector contains a large number of trench perpendicular normal faults, suggesting a stress rotation compared to the northern sector. We further examine the local stresses of the northern and southern sectors separately.
\newline 
\noindent
Interface activity in the northern sector shows widespread slab parallel thrusting and shallow dipping thrust events while we observe only a small active cluster in the south showing slightly different striking directions (Fig. \ref{fig:ResMap}). The stress inversion for 38 events results in azimuth and plunge of $\sigma_1$ consistent with the average slab geometry striking $\sim 63^\circ$ and dipping $\sim 29^\circ$ in these parts (Profile A Figure \ref{fig:XSection}). Stress inversion for the 15 events further south, however, shows a nearly slab-dip parallel compressional regime with maximum stress $\sigma_1$ at $88^\circ$ and a near-horizontal thrust angle $84^\circ$. A reduced R (0.32) indicates a more complex stress field compared to the northern part.
\newline 
\noindent
Upper plate seismicity is mainly normal faulting with a subset of strike-slip and NF/TF to SS transitional events. The stress inversion of 22 events in the northern crust reveals a transitional regime from strike-slip to normal faulting \citep{Lund2007} with $\sigma_1$ direction of 119◦. The small scale quantity R = 0.12 shows a large influence of $\sigma_2$ and suggests a strike-slip component. Further south we observe a rotated, arc-perpendicular stress regime with $\sigma_3$ trending $340^\circ$ and $\sigma_1$ plunging nearly vertically in a largely diffuse striking direction. The scale quantity indicates normal faulting dominates with R = 0.80, compared to the north. in a largely diffuse striking direction.
\newline 
\noindent
We perform a single stress inversion for the deep earthquake group in the southern sector. In this inversion,  we include 20 deep events, spatially clustered between Dominica and Martinique at $\sim$150 km depth. Contrasting to the stress condition on the interface at shallow depth, the deep events depict a slab-dip perpendicular extensional stress regime with $\sigma_1$ orientating parallel ($54^\circ$) to the slab and plunging steeply at $71^\circ$. The scale quantity R equals 0.69.

\section{Tectonic Implications}
\begin{figure}
\centering
\includegraphics[width=1.0\linewidth]{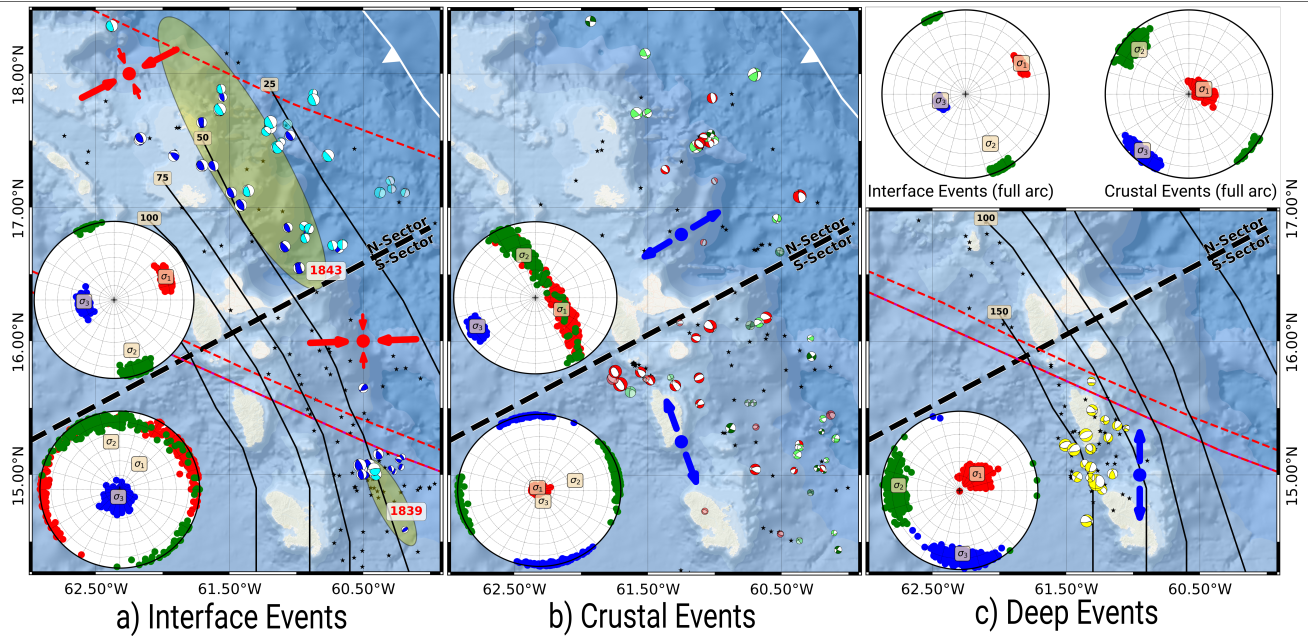}
\caption{Results of stress inversion of a) interface, b) crustal and c) deep events in the northern and southern sector. Top-right shows the inversion results of the interface and crust without regional sectoring.}
\label{fig:StressMap}
\end{figure}

\paragraph*{Interface activity}
At depths shallower than 50 km, we find a dominance of thrust activity and further towards the trench, shallow dipping thrust events become more frequent. This agrees with previous studies in the region \citep[e.g.][]{GONZALEZ2017214} and implies seismic slip along the megathrust between the subducted slab and overriding plate. In the northern sector (Fig. \ref{fig:StressMap} a), the stereo plot of the stress regime depicts widespread activity with the pattern of a classical subduction zone. Most of the events are located within the estimated rupture area of the 1843 M8.5 earthquake \citep{Feuillet2011b}. The occurrence of these thrust events corroborates that the 1843 event was a megathrust earthquake that ruptured the seismogenic part of the plate interface, instead of being within the subducted lithosphere \citep{vanRijsingen2020}. The southern sector lacks such compressional events except for in a small but highly active cluster (Fig. \ref{fig:ResMap}; Zoom Box 2) within the area of the 1839 M7.5 event to the east of Martinique which is also assumed to be an megathrust event \citep{Feuillet2011b}. The stereo plot for the southern sector indicates a much more diffuse pattern featuring a range of fault strikes, but still with an overall  trench-perpendicular compressional stress that appears to rotate along with the curve of the island arc (Fig. \ref{fig:StressMap} a). 

\begin{figure}
\centering
\includegraphics[width=1.0\linewidth]{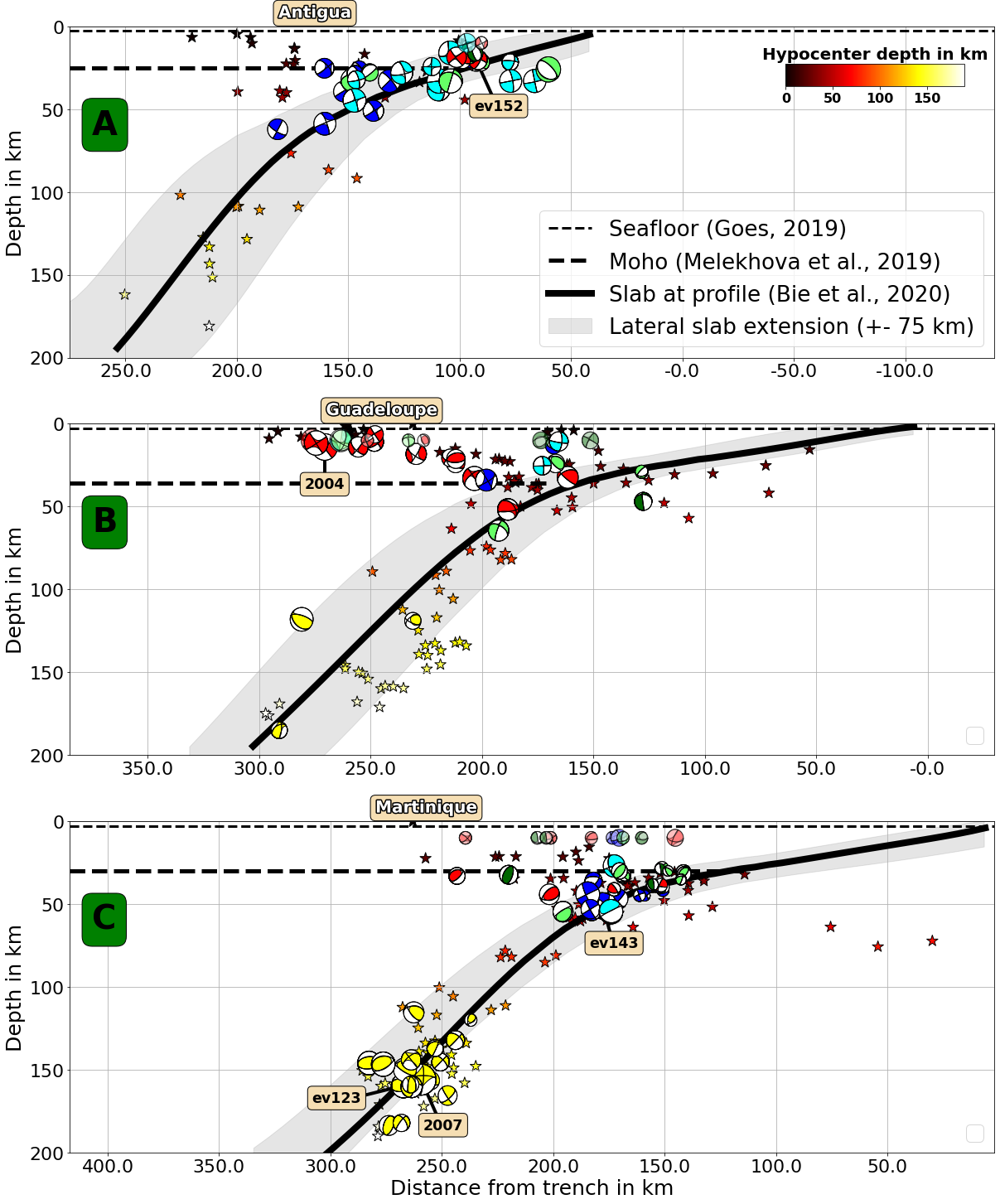}
\caption{Cross-section plots along depth profile A, B, and C. 
The featured seismicity and local slab model is derived by \cite{Bie2019}. Gray shaded area surrounding the slab curve at the profiles indicates the lateral extension to north and south ($\pm 75$ km).Thick dotted lines represent the average Moho depth at the profile intersection \citep{Melekhova2019} with the above thin dotted line being the average water depth of 2.8 km at which the OBS sensors are located \citep{Goes2019}. Color-coding of the displayed beachballs corresponds to the FMC-diagram (Fig. \ref{fig:FMC}). Faded beachballs do not have depth information and are set to a default depth of 10 km.}
\label{fig:XSection}
\end{figure}

\paragraph*{Upper plate activity}
Upper plate activity comprises a majority of normal faulting events but also includes a subset of SS and NF/TF to SS transitional mechanisms. Transitional events occur primarily in the northern sector, where our results (Fig. \ref{fig:StressMap} b) indicate a rotational stress regime \citep{Lund2007}. Such rotation may be related to the transition from a sinistral strike-slip system at the northern end of the LA subduction zone, towards an approximately arc perpendicular extensional faulting regime \citep{Julien1989,Laurencin2017}. In the midst of this area, we recognise a small cluster of events with a variety of mechanisms (Fig. \ref{fig:ResMap}; Zoom Box 1) centered at a depth of around 20 km, still within the Caribbean crust but near to the interface zone. The cluster is located above the projected position of the Barracuda Ridge after subduction, inferring a possible link between them. Seaward of the trench, the Barracuda Ridge rises 1800 m above the surrounding seafloor and is topped by a very thin sedimentary drape \citep{Patriat2011}. Previous work did suggest this feature would have an effect on upper plate deformation \citep{McMann1984,Laigle2013b}, as seen in our study. The stress regime of the upper plate evolves from the transitional regime of the northern sector towards a dominance of  arc-parallel extension in the south (Fig. \ref{fig:StressMap} b).  However, even though the overall stress tensor in the southern sector indicates arc-parallel extension, events in the vicinity of the 2004 M6.3 Les Saintes earthquake reflect arc-perpendicular extension consistent with the arc-parallel en-echelon faults identified by \citet{Feuillet2011a}. While interpretations for events to the east of Martinique are limited by the small event set and mostly missing depth information, we recognise on cross-section profile B (Fig. \ref{fig:XSection}; B) that the depths of upper plate normal faulting increase towards the east. The events with shallow depths west of and aligned with the arc correspond to arc parallel normal faulting of the Les Saintes earthquake swarm. Their shallow depths and overall smaller contribution to the total moment released in normal faulting in the southern segment, may be due to the relatively high crustal temperatures at and near the arc. Eastwards, in the forearc, azimuthal directions of the normal faults become mostly arc perpendicular as their depths increase towards the interface zone. This change in style is consistent with change in orientation of the faulting structures identified by active seismic imaging \citep{Feuillet2011a}, and indicates that the fore-arc is characterised by stresses associated with along-arc bending and oblique subduction \citep{Feuillet2002} while the arc and back-arc are dominated by trench-perpendicular extension as expected due to pull of the underlying slab, possibly exploiting structures formed during previous phases of back-arc opening \citep{Allen2019}.

\begin{figure}
\centering
\includegraphics[width=1.0\linewidth]{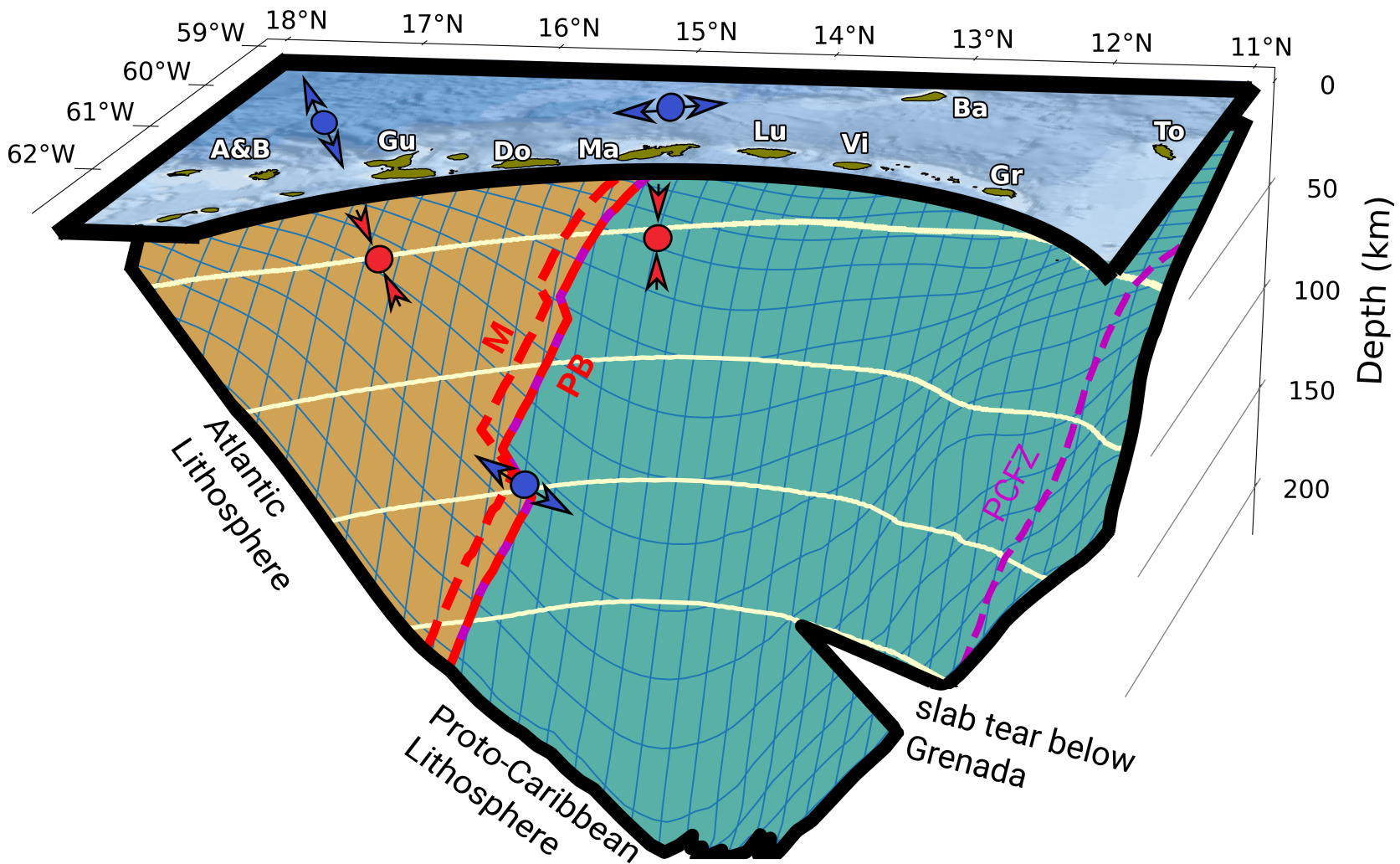}
\caption{3D sketch of the Lesser Antilles subduction zone. Marked islands from north to south are Antigua and Barbuda (\textbf{A\&B}), Guadalupe (\textbf{GU}), Dominica (\textbf{Do}), Martinique (\textbf{Ma}), St. Lucia (\textbf{Lu}), St. Vincent (\textbf{Vi}), Barbados (\textbf{Ba}), Grenada (\textbf{Gr}) and Tobago (\textbf{To}). Outline of the slab is based on the local slab model by \cite{Bie2019} with spacial partition into Atlantic (brown) and Proto-Caribbean (green) lithosphere \citep{Braszus2020}. Relevant features are displayed in color coding following Figure \ref{fig:Map1} with \textbf{M} being the Marathon fracture zone, \textbf{PB} the plate boundary between the two lithospheres which also coincides with the Mercurius fracture zone and the unnamed Proto-Caribbean fracture zone \textbf{PCFZ}. Displayed arrows represent the dominant stress regime, projected at the centroid location of the input data. The hypothesised lateral slab tear below Grenada is depicted at the depth outline of 200 km.}
\label{fig:3D_cartoon}
\end{figure}

\paragraph*{Slab activity}  
Seismic activity within the intermediate-depth range (70-300 km) concentrates between Dominica and Martinique. Here the comparably large size of our MT database (N=18) and the high quality event locations \citep{Bie2019} allow us to perform an examination of the activity that we attribute to intraslab seismicity. This cluster encloses several large magnitude events, including the 2007 Mw7.4 Martinique event and the Mw5.8 ev123 event of the VoiLA catalog (Tab.\ref{tab:RefEvents}). Both moment tensor solutions have a relatively large CLVD component suggesting a complex rupture history. In general, this cluster shows predominantly normal faulting activity with fault strikes in a trench perpendicular direction in the north (under Dominica) and trench parallel towards the south (between Dominica and Martinique). We also find a more variable set of mechanisms including some strike-slip and a single thrust mechanism in the vicinity of the 2007 Martinique event. The dominant stress condition is downdip compression (as $\sigma_1$ axes are most strongly clustered) and slab parallel extension (Fig. \ref{fig:StressMap} c). Most subduction zones exhibit downdip extension at intermediate depths, but a few without deep seismicity (below 300 km depth) display downdip compression within the 100-200km depth range, including New Britain, Ryukyu and the central Aleutians, as well as a few other segments of trenches with deep seismicity (central Tonga, northern Kuriles) \citep{Chen2004,Alpert2010,Bailey2012}.  Some further exceptions of this can be found in more complex settings like South Sandwich that exhibits a local stress reversal in depth below 150 km \citep{GinerRobles2009}, in Calabria with downdip compression between 200-300 km depth \citep{Bailey2012} or the Hellenic arc with a slab perpendicular compressional stress regime at its center \citep{Rontogianni2010}. 
\newline 
\noindent
Downdip compression at intermediate depths may be the result of a relative high resistance to sinking into the lower mantle \citep{Goes2017}, which would be consistent with the interpretation from \citet{Braszus2020}, that subducted slab material from 70 Myr ago to present has accumulated in the upper mantle below the Antilles. In most of the slabs with downdip compression at intermediate depths however, the extension is oriented slab-perpendicular, while in the Antilles, the extension is dominantly slab-parallel (Fig. \ref{fig:StressMap} c). Geometric effects of a curved slab are not sufficient to explain the rotational behavior of the Antillean intermediate-depth normal faults. Along-strike bending would introduce along-strike compression on the inside of the bend as well as along-strike extension on the outside of the bend. One might expect that the compression closer to the colder slab top might be more prominent in seismicity, yet we see only extension. Trench-parallel extension was found in the northern Hikurangi subduction zone, New Zealand \citep{Okuwaki2021}; however, this stress pattern was attributed to a subducting oceanic plateau or seamounts in a centroid depth of only around 70 km, which is not a likely mechanism for the LA. Wadati-Benioff seismic activity below central LA has long been hypothesized to be influenced by a gap in the slab near $15^\circ$N \citep{Bie2019}. Seismic anisotropy from teleseismic SKS waves \citep{Schlaphorst2017} was interpreted as due to mantle material flowing through a gap within the subducted lithosphere at around 170-200 km depth. A gap at this location was also supported by teleseismic tomography models \citep{Benthem2013, Harris2018} of the central parts of the arc ($\sim 15^\circ$N). A recent teleseismic tomography study, however, by \citet{Braszus2020} showed that any gap is confined to the deep upper mantle and the slab looks continuous at the depths of the seismicity. Instead \citet{Braszus2020} observes a wavespeed anomaly that can be attributed to a signature of excess slab hydration within the subducted Proto-Caribbean plate, possibly near the subducted spreading center. The seismicity does reveal a change in slab dip (Fig. \ref{fig:XSection}), with the northern part of the slab being steeper and deeper than the southern part, but the vertical offset is not sufficient to generate a gap in the slab \citep{Bie2019}. Furthermore, \cite{Bie2019} find no sharp changes in the slab dip angle that would be expected if there was a tear, but they do note a thickening of the Wadati–Benioff zone around $\sim$150 km. The lateral changes in deformation style revealed by our MT dataset do indicate that a structure within the subducted Proto-Caribbean plate may be present. The subducted domain boundary between plate material formed at the Proto-Caribbean (to the south) and Central Atlantic ridge (to the north) projects just north of the cluster (Fig. \ref{fig:ResMap}; magenta-red line), and part of the extinct Proto-Caribbean spreading ridge is expected to lie below the central arc  (Fig. \ref{fig:3D_cartoon}).
A lateral variation in slab buoyancy (more buoyant in the south than north) might explain the along-strike variation in slab dip and introduce along-strike extensional stresses. However, the age, and hence negative buoyancy of the subducted plate increases southwards \citep{Braszus2020} ($\sim 30 Myr older in the south)$, although all of the plate is older than 70 Myr and hence age-related buoyancy gradients may be modest. 
On the other hand, a partial detachment of the southern end of the slab, along a lateral tear recognised at around 200 km depth below Grenada \citep{Braszus2020}, may be the cause of weaker slab pull in the south than in the north. Pre-existing Proto-Caribbean structures in the downgoing plate may govern both the localisation of Wadati-Benioff seismicity in response to the stress regime due to along-strike slab pull variations, and the orientation of the seismically activated structures (Fig. \ref{fig:3D_cartoon}).

\section{Conclusions}
We have developed a novel Am$\Phi$B-RMT approach which employs the uniXtree-sampling algorithm to mitigate the effects on RMT solutions from OBS data of uncertainties from horizontal station alignments and high levels of ocean microseisms. Our synthetic experiments show a reduction of the discrepancies between input and recovered CLVD percentage to be between 48\% to 58\% compared to a classical L2 approach, depending on the number of stations used and the noise levels. We also observe a reduction in error in the recovered Kagan angle at large noise amplitudes, with up to 40\% improvement compared to a classical L2 approach. Inversions taking into account station alignment uncertainties at low noise are more robust if a model covariance matrix is included. Using only vertical traces it is possible and robust to solve in a double-couple inversion setup, as changes in CLVD are shown to mainly compensate for the effects of noise, but do not generally bias the retrieved planar fault geometry.
\newline
\noindent
Application of our new method to the recorded OBS data of the VoiLA experiment in the Lesser Antilles subduction zone yields robust solutions as shown by the comparison with a teleseismically-derived reference event set. We compiled 151 MT solutions, including 39 new ones from this study for the north-central Lesser Antilles arc, and found 25\% normal faulting (NF), 25\% thrust faulting (TF), 10\% strike-slip faulting (SS) with the remainder being a mixture of mainly NF to SS transitional mechanisms. TF is dominant along the slab-upper-plate interface, with a notable gap in seismic activity between $\sim15^\circ$N and $\sim16.5^\circ$N. The general stress regime follows the subduction geometry of the slab. We find that the areas of interface activity coincide with the inferred rupture areas of the 1839 M7.5 and 1843 M8.5 megathrust events. Upper plate activity shows a strong variability along the arc. The north is characterized by a SS to NF transitional regime while the stress condition further south is dominantly an arc parallel extensional regime probably due to along-arc stretching of the fore-arc to accommodate the tight arc curvature.
Stress inversion for an intermediate-depth central cluster offshore Martinique reflects down-dip compression and slab-parallel extension. Down-dip compression may reflect the dynamics of this relatively small subduction zone \citep{Benthem2010}, which has led to the piling of slab material within the upper mantle below the LA \citep{Braszus2020}. The along-strike extensional stress regime is unusual and we propose it to be due to lateral buoyancy variations, where pull at the southern end of the LA slab may be reduced by partial slab detachment along a lateral slab tear below Grenada \citep{Braszus2020}, causing slab deformation that localizes along pre-existing structures around the extinct Proto-Caribbean ridge subducted below Martinique (Fig. \ref{fig:3D_cartoon}).

\section*{Acknowledgments}

\section*{Data availability}
The data underlying this article are available on the IRIS web service, at \url{https://doi.org/10.7914/SN/XZ_2016}.

\section*{Code availability}
The moment tensor inversion code is available on request from Mike Lindner [Karlsruhe Institute of Technology]

\bibliography{BibTeX}
\bibliographystyle{apalike}


\newpage
\begin{table}
\caption{Table of reference events. The table features two different depth information (in order): the depth given by USGS and the depth (blue) derived by \cite{Bie2019}. Lateral information are similar in both sources, displayed are the location by \cite{Bie2019} which we also use in the inversion process. Entries marked in blue are findings of this study. Our beachballs are displayed for 95\% (black) and 97.5\% (red) probability solution.}
\begin{center}
\begin{tabular}{||c|c|c|c|c|c|c|c||} 
\hline\hline
ID & Day & lat | lon & $M_w$ & FM & CLVD & Lit. & This\\
& Onset-Time & depth | \textcolor{blue}{depth} & \textcolor{blue}{$M_w$} & \textcolor{blue}{FM} & \textcolor{blue}{CLVD} & USGS &
Study\\\hline\hline
ev106&2016-05-09&16.179 | -60.615&4.4&77,57,-115&-22& 
\multirow{2}{*}{\includegraphics[width=0.06\linewidth]{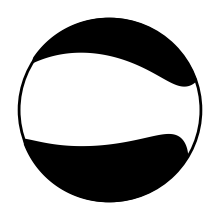}}&
\multirow{2}{*}{\includegraphics[width=0.06\linewidth]{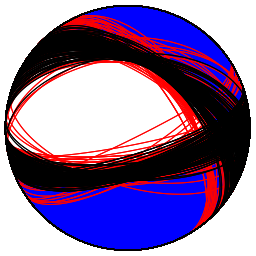}}\\
&13:36:27.76&23.9 | \textcolor{blue}{32.34}&\textcolor{blue}{4.7}&\textcolor{blue}{78,57,-114}&---&&\\\hline
ev122&2016-10-14&16.723 | -60.655&4.4&180,2,105&-4& 
\multirow{2}{*}{\includegraphics[width=0.06\linewidth]{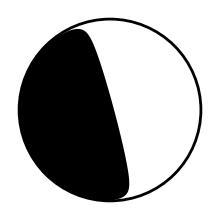}}&
\multirow{2}{*}{\includegraphics[width=0.06\linewidth]{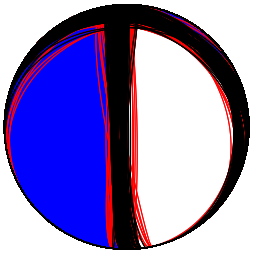}}\\
&17:25:24.00&25.3 | \textcolor{blue}{10.65}&\textcolor{blue}{4.7}&\textcolor{blue}{179,83,-79}&---&&\\\hline
ev123&2016-10-18&15.298 | -61.352&5.6&248,44,-106&2&
\multirow{2}{*}{\includegraphics[width=0.06\linewidth]{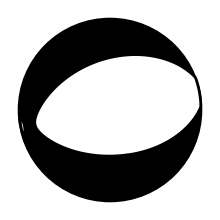}}&
\multirow{2}{*}{\includegraphics[width=0.06\linewidth]{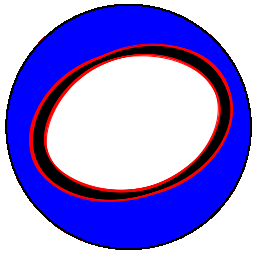}}\\
&22:07:43.75&146.0 | \textcolor{blue}{159.62}&\textcolor{blue}{5.6}&\textcolor{blue}{246,42,-93}&\textcolor{blue}{9}&&\\\hline
ev143&2017-02-03&15.065 | -60.457&5.80&334,59,86&1&
\multirow{2}{*}{\includegraphics[width=0.06\linewidth]{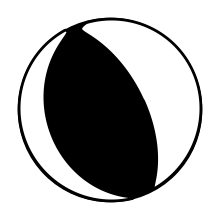}}&
\multirow{2}{*}{\includegraphics[width=0.06\linewidth]{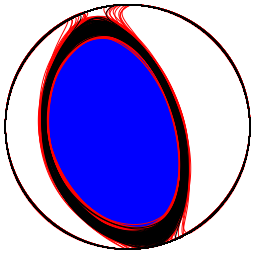}}\\
&19:53:52.86&44.0 | \textcolor{blue}{51.14}&\textcolor{blue}{5.9}&\textcolor{blue}{341,55,87}&\textcolor{blue}{-2}&&\\\hline
ev152&2017-04-17&17.513 | -61.025&4.8&133,37,-87&9&
\multirow{2}{*}{\includegraphics[width=0.06\linewidth]{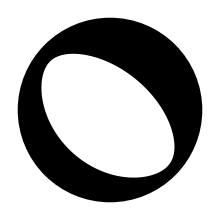}}&
\multirow{2}{*}{\includegraphics[width=0.06\linewidth]{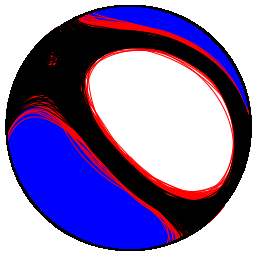}}\\
&06:24:41.40&18.7 | \textcolor{blue}{20.59}&\textcolor{blue}{5.0}&\textcolor{blue}{140,68,-70}&\textcolor{blue}{2}&&\\\hline
ev155&2017-04-25&16.835 | -60.915&4.4&336,79,82&11&
\multirow{2}{*}{\includegraphics[width=0.06\linewidth]{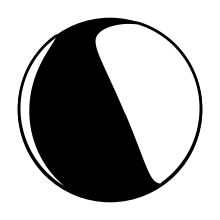}}&
\multirow{2}{*}{\includegraphics[width=0.06\linewidth]{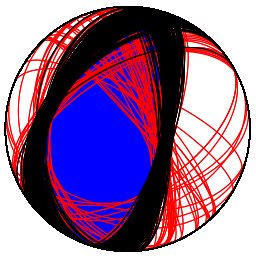}}\\
&09:53:31.81&38.5 | \textcolor{blue}{15.71}&\textcolor{blue}{4.5}&\textcolor{blue}{9,61,80}&---&&\\\hline\hline
\end{tabular}
\end{center}
\label{tab:RefEvents}
\end{table}

\begin{table}
\caption{Stress inversion}
\begin{center}
\begin{tabular}{||c|c|c|c|c|c|c||}
    \hline\hline
    Layer & Region & N & $\sigma_1$ azimuth in $\circ$/ & $\sigma_2$ azimuth in $\circ$/ & $\sigma_3$ azimuth in $\circ$/ & $R$ \\
    & & & plunge in $\circ$ & plunge in $\circ$ & plunge in $\circ$ &\\\hline\hline
    Interface & Arc & 53 & 61/25 & 153/3 & 248/65 & 0.52 \\\hline
    Interface & North & 38 & 63/29 & 157/8 & 262/60 & 0.55 \\\hline
    Interface & South & 15 & 88/84 & 359/9 & 150/80 & 0.32 \\\hline
    Upper plate & Arc & 73 & 75/80 & 274/10 & 183/3 & 0.87 \\\hline
    Upper plate & North & 22 & 119/55 & 340/28 & 239/19 & 0.12 \\\hline
    Upper plate & South & 51 & 234/88 & 70/2 & 340/1 & 0.80 \\\hline
    Slab & South & 18 & 54/71 & 273/15 & 180/11 & 0.69 \\\hline\hline
\end{tabular}
\end{center}
\label{tab:StressInv}
\end{table}

\end{document}